\documentclass[preprint,12pt]{elsarticle}
\usepackage{subfigure}
\usepackage{multirow}
\usepackage{graphicx}
\usepackage{amsmath}
\usepackage{natbib}%[square, numbers, comma, sort&compress]
\usepackage{lineno}
\usepackage[section,subsection,subsubsection]{extraplaceins}
\usepackage[colorlinks=true, linkcolor=blue, backref=page]{hyperref}

\journal{Computer Physics Communications}

\begin{document}
\begin{frontmatter}

\makeatletter
\def\@author#1{\g@addto@macro\elsauthors{\normalsize%
    \def\baselinestretch{1}%
    \upshape\authorsep#1\unskip\textsuperscript{%
      \ifx\@fnmark\@empty\else\unskip\sep\@fnmark\let\sep=,\fi
      \ifx\@corref\@empty\else\unskip\sep\@corref\let\sep=,\fi
      }%
    \def\authorsep{\unskip,\space}%
    \global\let\@fnmark\@empty
    \global\let\@corref\@empty  %% Added
    \global\let\sep\@empty}%
    \@eadauthor={#1}
}
\makeatother



\title{Error Propagation of the Track Model and Track Fitting Strategy for the Iron CALorimeter Detector in India-based Neutrino Observatory\tnoteref{titlelabel}}
\tnotetext[titlelabel]{This work is submitted on behalf of INO collaboration}

\author{Kolahal Bhattacharya\corref{cpa}}
\cortext[cpa]{Corresponding author}
\ead{kolahalb@tifr.res.in}

\author{Arnab K Pal}
%\ead{arnabkp@gmail.com}
\author{Gobinda Majumder}
%\ead{gobinda@tifr.res.in}
\author{Naba K Mondal}
%\ead{nkm@tifr.res.in}

% \fntext[akp]{left the collaboration}
% \fntext[nkm]{spokesperson}
% \fntext[gm]{code guru}
\address{Department of High Energy Physics, INO-Group (HECR Section), Tata Institute of Fundamental Research, 1-Homi Bhabha Road, Colaba, Mumbai-400005, India}

\begin{abstract}
A Kalman filter package has been developed for reconstructing muon 
$(\mu^\pm)$ tracks (coming from the neutrino interactions) in ICAL 
detector. Here, we describe the algorithm of muon track fitting, with 
emphasis on the error propagation of the elements of Kalman state 
vector along the muon trajectory through dense materials and inhomogeneous 
magnetic field. The higher order correction terms are included for 
reconstructing muon tracks at large zenith angle $\theta$ (measured 
from the perpendicular to the detector planes). The performances of 
this algorithm and its limitations are discussed.
\end{abstract}

\begin{keyword}
Kalman filter \sep track fitting \sep magnetized iron calorimeter\\
PACS: 07.05.Kf \sep 29.40.Vj \sep 29.40.Gx
\end{keyword}
\end{frontmatter}

%\linenumbers
\section{Introduction}\label{intro}
Fitting a charged particle's track through dense materials and inhomogeneous magnetic 
field is a well-known problem in High Energy Physics experiments. This is achieved by a 
recursive least square method, known as the Kalman filter~\cite{1960Kalman}. For linear 
systems, this filter gives the best possible estimate with no bias and minimum variance~\cite{9780521635486}. However, for non-linear systems (like track fitting), there is no 
optimal estimator. The principles of the Kalman filter are generalized for constructing 
an extended Kalman filter~\cite{fujiiextended} or an unscented Kalman filter~\cite{wan2000unscented}. 
Typically, an extended Kalman filter is used for track fitting by which one can extract 
the information of charge and momentum of the particle.

\begin{figure}[ht]
\centering
\subfigure[ICAL Geometry]
{
  \includegraphics[width=0.45\textwidth,height=0.35\textwidth]{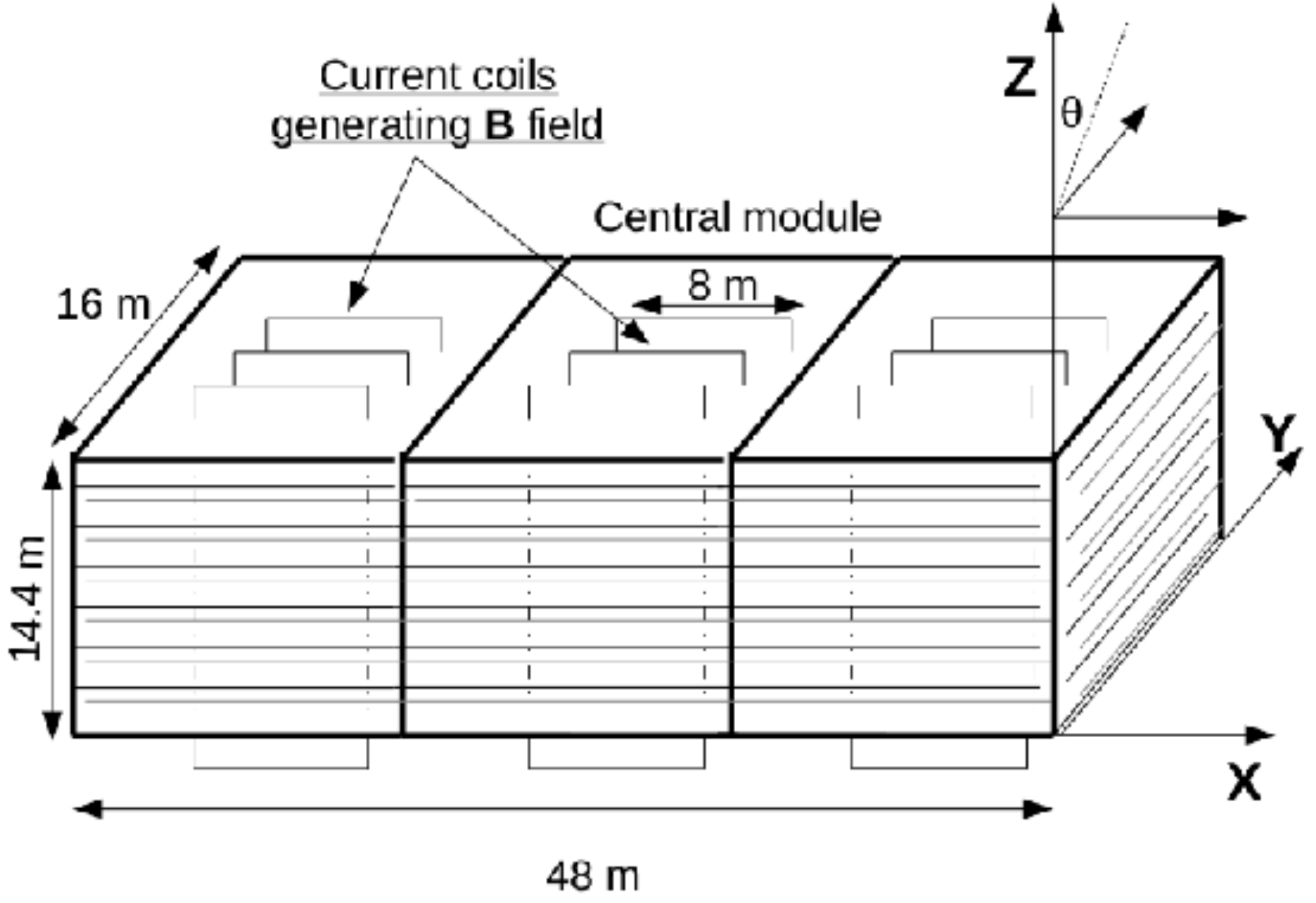}
  \label{f1(a)}
}
\hspace{0.05 cm}
\subfigure[$\bf{B}$ field map]
{
  \includegraphics[width=0.45\textwidth,height=0.35\textwidth]{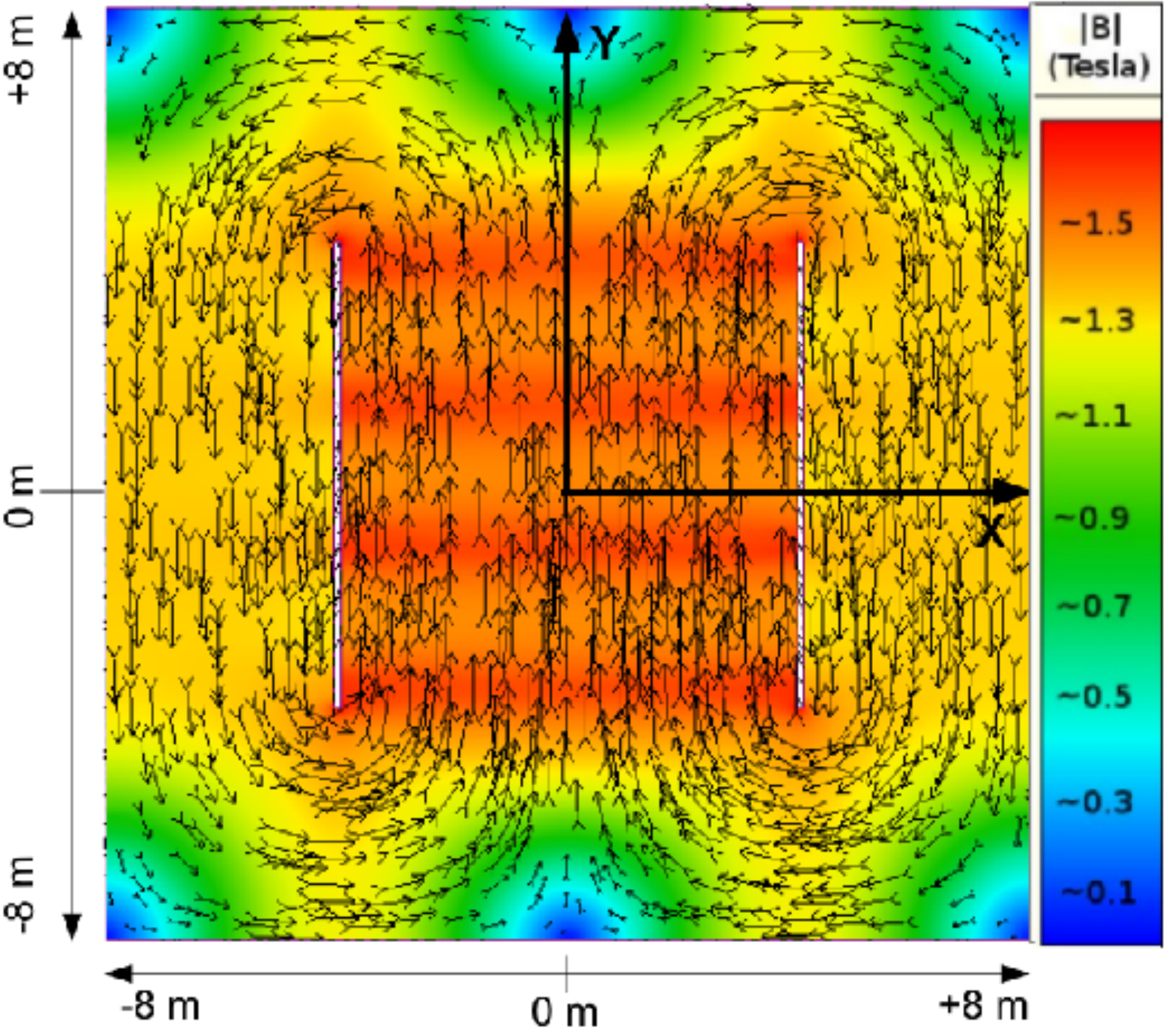}
  \label{f1(b)}
}
\caption{(a) ICAL detector geometry and (b) Magnetic field map shown in central 
module. The same field pattern exists in side modules as well.}
\end{figure}

The Iron CALorimeter (ICAL) detector proposed by India-based Neutrino Observatory (INO) 
Collaboration~\cite{athar2006report} will observe atmospheric neutrinos $(\nu)$ through 
a 50 kiloton magnetized iron tracking calorimeter. The detector comprises three modules, 
each of dimension $16\rm{m}\times 16\rm{m}\times14.4\rm{m}$ (Fig.~\ref{f1(a)}). These 
modules have 150 active detector planes, sandwiched between 151 iron slabs of thickness 
5.6 cm. Each active plane in a module is made up of sixty four (64) $2\rm{m}\times2\rm{m}
$ Resistive Plate Chamber (RPC) detectors~\cite{bheesette2009design}. The current coils 
(Fig.~\ref{f1(a)}) generate magnetic field (Fig.~\ref{f1(b)}) inside the iron slabs. 
Charge current (CC) $\nu$ interactions will produce muons, electrons and hadrons which 
will pass through the detector and will give signals at the RPC strips.%Outside the iron slabs, the field is zero. 

The simulation effort for ICAL experiment begins with coding the detector geometry with 
the GEANT4~\cite{Agostinelli2003250} tool-kit and by implementing the magnetic field map. 
The signals left by any particle in RPC detectors along its track are digitized to form 
hits in a rectangular ICAL coordinate system (Fig.~\ref{f1(a)}). The hits due to muon, 
a minimum ionizing particle, can be joined to form a track candidate. A track candidate 
needs to have minimum five hits in consecutive active layers to be defined as a muon track. 
The Kalman filter is used to fit this track.

On the other hand, hadron hits form a shower which might be calibrated with rather poor 
resolution~\cite{devi2013hadron}. Electrons also generate showers quickly and within 4-
5 iron layers they get absorbed. Therefore, it is not possible to reconstruct electrons. 
Thus, good reconstruction of muon is very important in this experiment. 

A Few previous attempts~\cite{Ghosh:2009ea, MuLkUpTbl} of designing a muon track fitting 
package were not very successful. Muon tracks with vertices smeared over a large volume 
of the detector (beyond the central $8\rm{m}\times8\rm{m}$ region of uniform magnetic 
field (Fig.~\ref{f1(b)})) were found prone to poor reconstruction. In reconstructed 
momentum distributions long tails were observed and charge identification efficiencies 
were poor as well. The aim of the current work is to design an algorithm with improved 
and stable performance of muon track fitting.

The main goal of the experiment is to resolve the neutrino mass hierarchy problem~\cite{9780198508717, smirnov2013neutrino}. 
It can also carry out the precision measurement analysis of atmospheric neutrino mixing 
parameters: $\Delta m^2_{32}$ and $\sin^2\theta_{23}$. So far, these analyses have been 
done using simple assumptions of detector response~\cite{Samanta:2006sj,Samanta:2007ue}. 
Recent studies~\cite{Ghosh:2012px,Thakore:2013xqa} were also done based on the detector 
performance of muon reconstruction, tabulated in the form of a look up table~\cite{MuLkUpTbl} 
that was generated by fitting Monte Carlo (MC) muon tracks with vertices smeared across 
the central $8\rm{m}\times8\rm{m}$ region of central module of the ICAL detector. The 
event-by-event analysis of the reconstructed data has not yet been performed. This work 
is of relevance from that aspect also.

We begin by reviewing the problem of track following with Kalman filter in section~\ref{TrackFollowing}. 
Next we explain the track following algorithm along the track in section~\ref{Algorithm}. 
In section~\ref{propagator}, we give the formulae for the error propagation of the track 
parameters. In section~\ref{results} the performance of the track following code and its 
limitations are presented and discussed.

\section{Kalman Filter}\label{TrackFollowing}

In the ICAL  experiment, the  active detector planes  are at  predefined $z$
coordinates. In such a case, the  state vector ${\bf x}=(x, y, t_x, t_y,
q/P)^T$ at each detector plane $z$  is very convenient to work with. The
elements  $x(z_k)$ and  $y(z_k)$ are  the coordinates  of a  hit at  the
$k^{th}\ z$ plane, expressed in  global ICAL coordinates. The charge $q$
of the particle  and its momentum $P$ at $k^{th}\  z$ plane are included
in the signed  inverse momentum element $q/P$.  The corresponding slopes
$t_x(z_k)$        and        $t_y(z_k)$       are        given        as
$t_x=dx/dz$      and     $t_y=dy/dz$.   %\left[\frac{dy}{dz}\right]
These  slopes  are  related to  the  zenith  angle
$\theta$  and   the  azimuthal  angle  $\phi$   through  the  relations:
$\cos\theta=\pm1/{\sqrt{1+t_x^2+t_y^2}}$ and $\tan\phi=t_y/t_x$ respectively.

To employ a Kalman filter, we must propagate the mean value of the state
$\bf{x}$  along with its associated errors~\cite{Fruhwirth:1987fm,
fontana2007track}. The filter routine then yields the near-optimum 
estimate  of  the  state  parameters. The  convergence  is  achieved  by
attributing measured importance to both the observed data as well as the
mathematical model giving  the prediction. The steps for  this are given
below.

\subsection{Basic Definitions of Kalman Filtering}

The equation  that describes  the evolution of  the state  $\bf{x}$ from
site $(k-1)$ to site $k$, is given by

\begin{equation}\label{2.1}
 {{\bf{\bar x_k}}}=f_{k-1}({\bf{\bar x_{k-1}}})+{\bf{w_{k-1}}}
\end{equation}
The bar symbol indicates the true values. Here $f_{k-1}$ is a propagator 
function that corresponds to a smooth deterministic motion in the absence 
of any random process noise $\bf w_{k-1}$. In a dense medium, the random 
noise comes from multiple Coulomb scattering and energy loss fluctuations~\cite{fontana2007track}. The process noise covariance matrix is given by 
$Q_k=\langle\bf{w_k}\bf{w_k}^T\rangle$. With an estimate $\bf{x_{k-1}}$ 
for any true state $\bf{\bar{x}_{k-1}}$, the estimation error covariance 
matrix is given by~\cite{fujiiextended}:

\begin{equation}
 C_{k-1}=\langle({\bf{x_{k-1}}}-{\bf{\bar{x}_{k-1}}})({\bf{x_{k-1}}}-{\bf{\bar{x}_{k-1}}})^T\rangle
\end{equation}
The detector  measures the state ${\bf  m_k}=(x(z_k),y(z_k))^T$ at every
plane ($z_k$)  along the particle  trajectory. The relation  between the
true state  vector $\bf{\bar{x}_k}$ and  the measured state  at $k^{th}$
plane is given by the measurement equation:

\begin{equation}\label{2.2}
{\bf{m_k}}=h_k({\bf{\bar x_k}}) + {\bf{\epsilon_k}}
\end{equation}
where ${\bf m_k}$ is the measured data, $h_k$ is the measurement function 
and $\bf{\epsilon_k}$ denotes the error associated with the measurement. 
As before, the measurement noise covariance is given as $V_k=\langle{\bf
{\epsilon_k}}{\bf{\epsilon_k}}^T\rangle$. It is assumed that the process 
noise and the measurement noise are Gaussian errors and $\langle Q_k\rangle=\langle V_k\rangle=0$. 

\subsection{Prediction}\label{predicting}

The latest state estimate $\bf{x}$ is extrapolated from $(k-1)^{th}$ 
plane to $k^{th}$ plane in the absence of random noise:

\begin{equation}\label{2.3}
 {{\bf{x_k^{k-1}}}}=f_{k-1}({\bf{x_{k-1}}})
\end{equation}
Here, $\bf{x_k^{k-1}}$ is the predicted state at $k^{th}$ plane from the measurements done up to 
$(k-1)^{th}$ plane. 
%Those packages that propagate the mean values of the state parameters usually employ Runge Kutta 4 technique or helix model. 
The function $f_{k-1}$ is non-linear in general. To propagate the mean value of the state vector, 
Runge Kutta 4 method or equations based on helix model~\cite{fujiiextended} are used in tracking. 
More recent work was done by Gorbunov et.$al$, who deduced an iterative analytic solution for $f
_{k-1}$~\cite{Gorbunov:2006pe}. We used this solution in our work. The propagation of the ($5\times5$) 
error covariance matrix $C$ is done as:

\begin{equation}\label{2.4}
 C_k^{k-1}=F_{k-1}\ C_{k-1}\ F_{k-1}^T + Q_{k-1}
\end{equation}
Here, ${C_k^{k-1}}$ is the predicted error matrix at $k^{th}$ plane from the measurements done up to 
$(k-1)^{th}$ plane. The derivation of Eq.\eqref{2.4} from Eq.\eqref{2.1} requires us to define 
$F_{k-1}$ as the $5\times5$ Jacobian matrix:

\begin{equation}\label{2.5}
F_{k-1}=\frac{\partial f_{k-1}}{\partial \bf x_{k-1}}
\end{equation}
The first term in Eq.\eqref{2.4} propagates the errors in the state from the $(k-1)^{th}$ plane 
to the $k^{th}$ plane. This is deterministic error propagation that comes from the magnetic field and the 
local slopes and momenta of the track. The ${5\times5}$ process noise matrix term $Q_k$ adds the 
random errors (due to multiple scattering and energy loss fluctuations) to the total error.

\subsection{Filtering}\label{filtering}

The next step is to minimize the incremental $\chi^2$ between $(k-1)^{th} $ plane 
to $k^{th}$ plane. The incremental $\Delta\chi^2$ is:

\begin{equation}\label{2.6}
\Delta\chi^2=({\bf{x_k}}-{\bf{x_k^{k-1}}}){\left[C_k^{k-1}\right]}^{-1}({\bf{x_k}}-{\bf{x_k^{k-1}}})^T + ({\bf{m_k}}-h_k({\bf{x_k}})){\left[V_k\right]}^{-1}({\bf{m_k}}-h_k({\bf{x_k}}))^T
\end{equation}
The term ${\bf{x_k}}$ denotes the Kalman estimate with respect to which the $\chi^2$ should be 
minimized. The condition $\frac{\partial}{\partial{\bf{x_k}}}(\Delta\chi^2)=0$ leads to Kalman 
estimate for state $\bf{x_k}$ at $k^{th}$ plane in terms of $(5\times2)$ Kalman Gain matrix $K 
_k$:

\begin{equation}\label{2.7}
 K_k=C_k^{k-1}\ H_k^T\left(H_k\ C_k^{k-1}\ H_k^T+V_k\right)^{-1}
\end{equation}
where the projector matrix $H_k$ is a $(2\times5)$ matrix given by $H_k=\frac{\partial\bf{m_k}} 
{\partial{\bf x_k^{k-1}}}$. For ICAL experiment, the projector matrix has the following form:

\[ \left( \begin{array}{ccccc}
1 & 0 & 0 & 0 & 0\\
0 & 1 & 0 & 0 & 0\end{array} \right)\]
and the $(2\times2)$ measurement error matrix $V_k$ is:

\[ \left( \begin{array}{cc}
{\sigma_x}^2 & 0\\
0 & {\sigma_y}^2\end{array} \right)\]
where $\sigma_x=\sigma_y=\frac{d}{\sqrt{12}}$, $d$ being the strip-width of RPC detector. 
In terms of Kalman Gain $K_k$, the filtered state estimate for $k^{th}$ plane is:

\begin{equation}\label{2.8}
 {\bf{x_k}}={\bf{x_k^{k-1}}} + K_k({\bf{m_k}}-H_k{\bf{x_k^{k-1}}})
\end{equation}
Similarly, the Kalman estimate for filtered error covariance $C_k$ is given by:

\begin{align}\label{2.9}
C_k &= (I-K_kH_k)C_k^{k-1}\nonumber\\
    &= (I-K_kH_k)C_k^{k-1}(I-K_kH_k)^T + K_kV_k{K_k}^T
\end{align}
Then, ${\bf x_k}$ and $C_k$ are used for extrapolation from $k^{th}$ plane to $(k+1)^{th}$ plane.
We found Eq.\eqref{2.9} (Joseph's form~\cite{9783540327950}) to be useful
for our work.

\section{Track Following}\label{Algorithm}

The track following algorithm is based on equations stated in section
~\ref{TrackFollowing}.
The coordinates of the digitized `hit' found closest to the vertex is 
used to initialize $(x, y)$ elements of the Kalman filter. Using hits 
in the first few planes, $(t_{x}, t_{y})$ near the vertex is obtained 
and are used to initialize the corresponding elements. The element $q
/P$ is initialized to zero, thereby nullifying any initial bias.

By taking steps $\Delta{z}$ in $z$ direction, the state vector ${\bf{
x}}(z+\Delta{z})$ is predicted from the state ${\bf{x}}(z)$. In every 
step, first the local material and magnetic field is found; then, the 
step size is decided. Inside iron layers, $\Delta{z}$ is set to 1 mm
and in other materials $\Delta{z}$ is set according to their widths. 
Energy loss of the particle in a step is calculated using Bethe Bloch 
formula~\cite{Nakamura:2010zzi} along with density effect corrections. 
The prediction of the $q/P$ element of the state vector is done using 
the energy loss information in that material. Prediction of the other 
elements are performed using the formulae given in section~\ref{KslF}.  
This way, the prediction of the state vector (Eq.\eqref{2.3}) is done. 
Once the current step is over, the next step uses the predicted state 
as the initial state for that step.

\begin{figure}[h]
\begin{center}
\scalebox{0.4}{\includegraphics{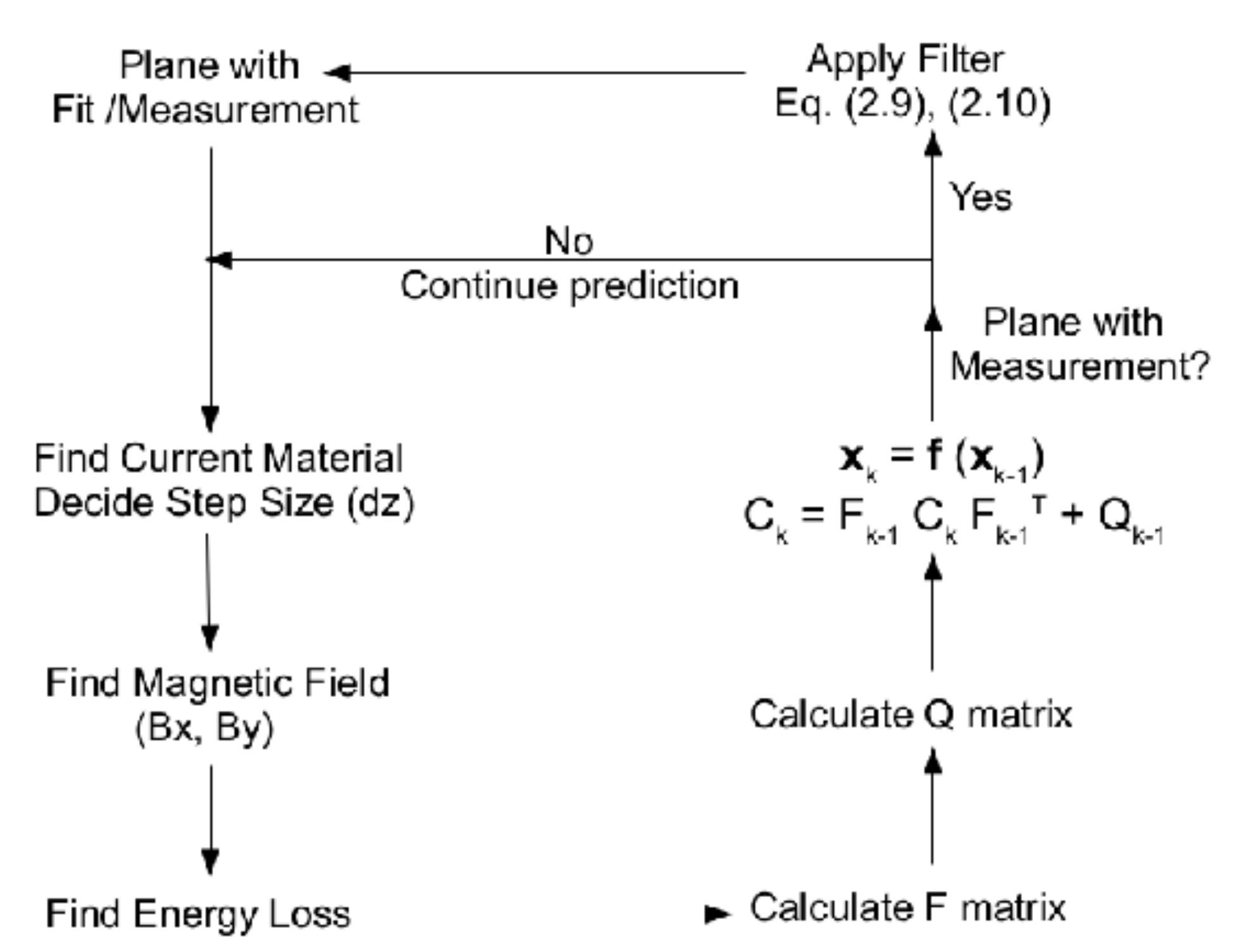}}
\caption{\label{f:algo} Schematics of the algorithm}
\end{center}
\end{figure}
In every step, the propagator matrix (Eq.\eqref{2.5}) and the random 
noise matrix are calculated locally, using the material and the 
magnetic field. They are used to propagate the errors associated with 
the state elements. Now, as tracking is done from one hit to the next 
through a series of thick and dense materials, Eq.\eqref{2.4} cannot 
be directly used between two hits (previous work~\cite{MuLkUpTbl} was 
based on direct use of Eq.\eqref{2.4} between two hits). It has to be 
used repeatedly in each successive steps so that starting from $(k-1)
^{th}$ hit, the random error contribution to the total error at $k^{t
h}$ hit becomes (Eq.(3.16) in~\cite{fujiiextended}):

\begin{equation}\label{2.10}
  Q_{k-1}=\sum_{s=1}^{N-1}F_{m_s,k}Q_{m_s}F_{m_s,k}^T
\end{equation}
In Eq.\eqref{2.10}, $F_{m_s,k}$ and $Q_{m_s}$ are the Jacobian matrix 
and the noise matrix for a small step. Then, the Kalman gain matrix $K
_k$ (Eq.\eqref{2.7}) is calculated from $C_k^{k-1}, V_k$ and $H_k$. It 
is used to for obtaining filtered state [from Eq.\eqref{2.8}] and updated 
error covariance matrix [from Eq.\eqref{2.9}].

After all the hits in the muon track candidate have been filtered, the 
hits are processed in the reverse order using the same algorithm. This 
procedure `smooths' the fitted track. The processing of hits in the 
forward and backward directions, completes one iteration. We have used 
four iterations, though for $>90$\% cases, the $\it fractional$ change 
in the desired state vector estimate were seen to become $<10$\% after 
the 2nd iteration. For tracks with only $4-5$ hits convergence was not 
at all achieved after four iterations.

\section{Error propagation of track model}\label{propagator}

The set $(x, y, t_x, t_y)$ and the corresponding errors $(\delta{x},\delta{y},\delta{t_x},\delta
{t_y})$ have been extrapolated on the basis of~\cite{Gorbunov:2006pe}. In this section, we shall 
give the formulae for error propagation of state parameters, to be used in the following propagator:

\begin{equation}\label{3.0}
F_{k-1}=
\begin{bmatrix}
 \frac{\delta [x(z_e)]}{\delta x(z_o)} & \frac{\delta [x(z_e)]}{\delta y(z_o)} & \frac{\delta [x(z_e)]}{\delta t_x(z_o)} & \frac{\delta [x(z_e)]}{\delta t_y(z_o)} & \frac{\delta [x(z_e)]}{{\delta(\frac{q}{P}}(z_o))}\\
 \frac{\delta [y(z_e)]}{\delta x(z_o)} & \frac{\delta [y(z_e)]}{\delta y(z_o)} & \frac{\delta [y(z_e)]}{\delta t_x(z_o)} & \frac{\delta [y(z_e)]}{\delta t_y(z_o)} & \frac{\delta [y(z_e)]}{{\delta(\frac{q}{P}}(z_o))}\\
 \frac{\delta [t_x(z_e)]}{\delta x(z_o)} & \frac{\delta [t_x(z_e)]}{\delta y(z_o)} & \frac{\delta [t_x(z_e)]}{\delta t_x(z_o)} & \frac{\delta [t_x(z_e)]}{\delta t_y(z_o)} & \frac{\delta [t_x(z_e)]}{{\delta(\frac{q}{P}}(z_o))}\\
 \frac{\delta [t_y(z_e)]}{\delta x(z_o)} & \frac{\delta [t_y(z_e)]}{\delta y(z_o)} & \frac{\delta [t_y(z_e)]}{\delta t_x(z_o)} & \frac{\delta [t_y(z_e)]}{\delta t_y(z_o)} & \frac{\delta [t_y(z_e)]}{{\delta(\frac{q}{P}}(z_o))}\\
 \frac{\delta [q/P(z_e)]}{\delta x(z_o)} & \frac{\delta [q/P(z_e)]}{\delta y(z_o)} & \frac{\delta [q/P(z_e)]}{\delta t_x(z_o)} & \frac{\delta [q/P(z_e)]}{\delta t_y(z_o)} & \frac{\delta [q/P(z_e)]}{{\delta(\frac{q}{P}}(z_o))}
\end{bmatrix}
\end{equation}

In this matrix the suffix $e$ refers to the Extrapolated position (${\bf{r}}+{d\bf{r}}$) while 
the suffix $o$ refers to the Old position ${\bf{r}}$. The last row is calculated more carefully 
for better convergence of $q/P$ parameter.

\subsection{State and error propagation for $(x, y, t_x, t_y)$}\label{KslF}
We have calculated the expressions for $x(z_e)$, $y(z_e)$, $t_x(z_e)$ and $t_y(z_e)$ in terms of 
$\left[x, y, t_x, t_y, q/P\right](z_0)$ and the magnetic field integrals, according to the model 
outlined in~\cite{Gorbunov:2006pe}. The analytic solutions were calculated up to the third order 
($n=2$ in $(q/P)^n$ with $n=0,1,2,3...$). They were utilized to find the derivatives required by 
the propagator matrix. 

\begin{align}\label{3.1}
x(z_e) &= x(z_0)+t_xdz+h\left(t_xt_yS_x-(1+t_x^2)S_y\right) + h^2[t_x(3t_y^2+1)S_{xx}\nonumber\\
       &- t_y(3t_x^2+1)S_{xy}-t_y(3t_x^2+1)S_{yx}+t_x(3t_x^2+3)S_{yy}]
\end{align}
\begin{align}\label{3.2}
y(z_e) &= y(z_0)+t_ydz+h\left((1+t_y^2)S_x-t_xt_yS_y\right) + h^2[t_y(3t_y^2+3)S_{xx}\nonumber\\
       &- t_x(3t_y^2+1)S_{xy}-t_x(3t_y^2+1)S_{yx}+t_y(3t_x^2+1)S_{yy}]
\end{align}
\begin{align}\label{3.3}
t_x(z_e) &= t_x+h\left(t_xt_yR_x-(1+t_x^2)R_y\right) + h^2[t_x(3t_y^2+1)R_{xx}\nonumber\\
         &- t_y(3t_x^2+1)R_{xy}-t_y(3t_x^2+1)R_{yx}+t_x(3t_x^2+3)R_{yy}]
\end{align}
\begin{align}\label{3.4}
t_y(z_e) &= t_y+h\left((1+t_y^2)R_x-t_xt_yR_y\right) + h^2[t_y(3t_y^2+3)R_{xx}\nonumber\\
         &- t_x(3t_y^2+1)R_{xy}-t_x(3t_y^2+1)R_{yx}+t_y(3t_x^2+1)R_{yy}]
\end{align}

In the above expressions, $h=\kappa(q/P)\sqrt{1+t_x^2+t_y^2}$ where $\kappa=0.29979$ $GeVc^{-1}T^{-1}
m^{-1}$ and $t_x$ and $t_y$ refer to $t_x(z_0)$ and $t_y(z_0)$ respectively. The factors $S_{...}$ and 
$R_{...}$ denote magnetic field integrals and $dz$ denotes $(z_e-z_0)$. The $B_z$ component of the ICAL 
magnetic field is zero and the field is in the $xy$ direction: $\vec{\bf{B}}=B_x(x(z),y(z)){\bf{\hat{x}}}+
B_y(x(z),y(z)){\bf\hat y}$. %The field is more or less homogeneous across the most of the detector volume, except for the corners and the area containing the magnetic coil. 
The field integrals are defined as 
~\cite{Gorbunov:2006pe}:
\begin{equation}\label{3.5}
 S_{i_1...i_k}=\int_{z_0}^{z_e}\int_{z_0}^{z_e}B_{i_1}(x(z_1), y(z_1))...\int_{z_0}^{z_e}B_{i_k}(x(z_k), y(z_k))dz_k...dz_1dz
\end{equation}
and 
\begin{equation}\label{3.6}
 R_{i_1...i_k}=\int_{z_0}^{z_e}B_{i_1}(x(z_1), y(z_1))...\int_{z_0}^{z_e}B_{i_k}(x(z_k), y(z_k))dz_k...dz_1
\end{equation}
where $i_1, i_2 ...$ etc denote $x, y, xx$ etc. These integrals were evaluated along the 
approximate particle trajectory. If the step size within iron is made reasonably small, 
magnetic field may be assumed to be constant along the step $dz$ and the calculation of 
the integrals becomes easier. 
\begin{table}[ht]
\caption{Magnetic Field Integrals}\label{Table1}
                                   % title of Table
\centering                         % used for centering table
\begin{tabular}{c c c c c c}       % centered columns (6 columns)
\hline\hline                       %inserts double horizontal lines
$S_x$ & $S_y$ & $S_{xx}$ & $S_{xy}$ & $S_{yx}$ & $S_{yy}$\\ [1.0ex]
                                   % inserts table
                                   %heading
%\hline                             % inserts single horizontal line
$\frac{1}{2}B_xdz^2$ & $\frac{1}{2}B_ydz^2$ & $\frac{1}{6}B_x^2dz^3$ & $\frac{1}{6}B_xB_ydz^3$ & $\frac{1}{6}B_xB_ydz^3$ & $\frac{1}{6}B_y^2dz^3$\\[1ex]
                                   % inserting body of the table
                                   %[1ex] adds vertical space
%\hline                             %inserts single line
\hline\hline                       %inserts double horizontal lines
$R_x$ & $R_y$ & $R_{xx}$ & $R_{xy}$ & $R_{yx}$ & $R_{yy}$\\ [1.0ex]
%\hline                             % inserts single horizontal line
$B_xdz$ & $B_ydz$ & $\frac{1}{2}B_x^2dz^2$ & $\frac{1}{2}B_xB_ydz^2$ & $\frac{1}{2}B_xB_ydz^2$ & $\frac{1}{2}B_y^2dz^2$\\[1ex]
\hline                             % inserts single horizontal line
\end{tabular}
\label{table:a}                    % is used to refer this table in the text
\end{table}
The field integrals $S_{...}$ and $R_{...}$ were evaluated assuming that $B_i(x(z), y(z))$ 
vary very slowly along the track $(x_{particle}(z),y_{particle}(z))$ and may be assumed to 
be constant when integrating with respect to $z$. This is true unless the particle is travelling 
almost parallel to the detector plane $(\theta\approx90^{o})$. The field integrals are given 
in Table~\ref{Table1}. The effect of fringe field just outside the iron layer has been neglected 
in this work.

We also included the transverse variation of the field as in~\cite{Wittek:931170} (first addendum). This 
is because an error in the position $(x,y)$ leads to an error in the magnetic field. 
For example, error in $B_x$ is:

\begin{equation}\label{3.7}
 \delta{B_x}\approx\frac{\partial B_x}{\partial x}\delta x + \frac{\partial B_x}{\partial y}\delta y
\end{equation}
The same is true for $B_y$ as well. This error in magnetic field gives an additional error of the 
direction of the track. As a result, there is an error $\delta{R_x}$ in the integral $R_x$:

\begin{align}\label{3.8}
\delta{R_x} &= \int\delta{B_x(x(z),y(z))\ dz}\nonumber\\
            &\approx\left[\frac{\partial B_x}{\partial x}\ dz\right]\delta{x}+\left[\frac{\partial B_x}{\partial y}\ dz\right]\delta{y}
\end{align}
Hence, from Eq.\eqref{3.3}, the error of $t_x(z_e)$ (to the first order in $h$) is given as:

\begin{align}\label{3.9}
\delta t_x(z_e) &= h\left[t_xt_y\frac{\partial B_x}{\partial x}-(1+{t_x}^2)\frac{\partial B_y}{\partial x}\right]dz\ \delta x \nonumber\\
                &+ h\left[t_xt_y\frac{\partial B_x}{\partial y}-(1+{t_x}^2)\frac{\partial B_y}{\partial y}\right]dz\ \delta y \nonumber\\
                &+ \left[1+h\left(t_y(1+\frac{{t_x}^2}{T^2})R_x-t_x(2+\frac{1+{t_x}^2}{T^2})R_y\right)\right]\delta t_x \nonumber\\
                &+ h\left[t_x\left(1+\frac{{t_x}^2}{T^2}\right)R_x-t_y\left(\frac{1+{t_x}^2}{T^2}\right)R_y\right]\delta t_y \nonumber\\
                &+ kT\ [t_xt_yR_x-(1+{t_x}^2)R_y]\ \delta(\frac{q}{P})
\end{align}
where $T=\sqrt{1+{t_x}^2+{t_y}^2}$. Similarly, smooth deterministic errors in other parameters can 
also be evaluated. Then, it becomes a trivial task to obtain the first four rows of the propagator 
matrix Eq.\eqref{3.0}. For instance, the term $\frac{\delta [t_x(z_e)]}{{\delta(\frac{q}{P}}(z_o))}$ is 
equal to $kT\ [t_xt_yR_x-(1+{t_x}^2)R_y]$ (from Eq.\eqref{3.9}). Terms of the order of $h^2$ were 
calculated using Mathematica~\cite{mathematica2011inc}.

\subsection{Signed inverse momentum}
\label{qperrorprop}
Signed inverse momentum element $q/P$ has been extrapolated with Bethe-Bloch energy loss formula~\cite{9783540572800}. 
The corresponding error propagation has been done using techniques shown in EMC internal reports~\cite{Wittek:931172} (second addendum). However, the tracks have been assumed orthogonal to the 
detector planes there. This is not true in INO-ICAL detector which would observe the atmospheric 
neutrinos coming from all directions. So, the error propagation of $q/P$ is done more rigorously.

We want to find out the error $\delta(q/P)$ at a point $(x(z+dz), y(z+dz), (z+dz))$ of the track 
in terms of the error $\delta(q/P)$ at $(x(z), y(z), z)$. We can write:

\begin{equation}\label{3.10}
 \delta(q/P)_{{{\bf{r}}(z+dz)}}=\delta(q/P)_{{\bf{r}}(z)}+\delta\left[d(q/P)\right]
\end{equation}
The first term in Eq.\eqref{3.10} refers to the error in the estimate of $(q/P)$, which was already 
there from $((x(z), y(z), z))$. The second term denotes the average systematic error that creeps 
in due to the incorrect estimation of $q/P$ at $(x(z+dz), y(z+dz), (z+dz))$ from that at $((x(z), 
y(z), z))$. It is possible to expand $d(q/P)$ as:

\begin{align}\label{3.11}
 d(q/P)&=(q/P)_{{{\bf{r}}(z+dz)}} - (q/P)_{{\bf{r}}(z)}\nonumber\\%|_{\bf{r}}^{{\bf{r}}+d{\bf{r}}} 
       &=f({{\bf{r}}(z+dz)}) - f({\bf{r}}(z))
\end{align}
where $f({\bf{r}})=q/P({\bf r})$ (which is known as the range-momentum relation). Muon CSDA 
(Continuous Slowing Down Approximation) range in $iron$ as function of muon momenta $P$ is known 
in the form of a numerical table~\cite{Groom:2001kq}. We evaluate $d(q/P)$ in the track frame such that 
no cross term arises in the following expansion:

\begin{equation}\label{3.12}
 d(q/P)=f'(l)dl+\frac{1}{2}f''(l)dl^2+\frac{1}{6}f'''(l)dl^3+...O(^4)
\end{equation}
where the arc length along the track is denoted by $l$. In this equation, it is assumed that the 
higher order correction terms are negligibly small. This approximation does not hold good if the 
particle track is at a large zenith angle $\theta>60^o$. In such cases, the derivatives of $f(l)
$ are small, of course; however, the factors containing $dl$ and its exponents grow rapidly as $
|dl|\approx\frac{dz}{\cos\theta}$. The error in $d(q/P)$ can be given by (from Eq.\eqref{3.12}):

\begin{align}\label{3.13}
 \delta\left[d(q/P)\right]&=\delta\left\lbrace f'(l)dl\right\rbrace+\delta\left\lbrace\frac{1}{2}f''(l)dl^2\right\rbrace+...\nonumber\\
                          &=f''(l)\delta{l}\ dl+f'(l)\ \delta(dl)+\frac{1}{2}f'''(l)\delta{l}\ dl^2+f''(l)\ dl\ \delta(dl)\nonumber\\
                          &=\left\lbrace f''(l)\ dl + \frac{1}{2}f'''(l)\ dl^2\right\rbrace\delta{l}+\left[f'(l)+f''(l)\ dl\right]\delta(dl)
\end{align}
One must find $\delta{l}$ and $\delta(dl)$ correctly to obtain the error $\delta(q/P)_{{\bf{r}}(z+dz)}$ as 
a function of $\delta(q/P)_{{\bf{r}}(z)}$ and others. The factor $\delta{l}$ may be found from:

\begin{align}\label{3.14}
 \delta(q/P)_l &=\delta{f(l)}\nonumber\\
               &=f'(l)\delta{l}
\end{align}
Thus, we have:

\begin{equation}\label{3.15}
 \delta{l}=\frac{\delta(q/P)_l}{f'(l)}
\end{equation}
The other term $\delta(dl)$ cannot be taken directly from EMC report $80/15$ as that calculation 
was done in SC frame $(x_\perp, y_\perp, z_\perp)$~\cite{fontana2007track, Wittek:931170} and we are 
working in a Cartesian reference frame. In Appendix, we show that the following holds in a Cartesian 
coordinate system [Eq.\eqref{6.6}]:

\begin{equation}\label{3.16}
\begin{bmatrix}
        \delta{x} \\
        \delta{y} \\
        \delta{z}
\end{bmatrix}_{{\bf{r}}(z+dz)}
=
\begin{bmatrix}
        1 & d\phi & -\cos\phi\ d\theta \\
       -d\phi & 1 & -\sin\phi\ d\theta \\
        \cos\phi\ d\theta & \sin\phi\ d\theta & 1
\end{bmatrix}
\begin{bmatrix}
        \delta{x} \\
        \delta{y} \\
        \delta{z}
\end{bmatrix}_{{\bf{r}}(z)}
+
\begin{bmatrix}
        \delta(dx) \\
        \delta(dy) \\
        \delta(dz)
\end{bmatrix}\,
%\label{eq:symmetrical}
\end{equation}
in a Cartesian frame, where $\theta$ and $\phi$ are the zenith and azimuthal angles respectively.
For ICAL experiment $\delta{z}_z=\delta{z}_{z+dz}=0$, as the detector planes correspond to fixed 
$z$ coordinates. Thus, from Eq.\eqref{3.16}, $\delta(dz)=-\cos\phi\ d\theta\ (\delta{x})_z-\sin
\phi\ d\theta\ (\delta{y})_z$. Then, $\delta(dl)$ may be expressed as:

\begin{align}\label{3.17}
 \delta(dl)&=\delta\left({dz\sqrt{1+{t_x}^2+{t_y}^2}}\right)\nonumber\\
           &=T\delta(dz)+dz\left(\frac{t_x}{T}(\delta\ t_x)_z+\frac{t_y}{T}(\delta\ t_y)_z\right)\nonumber\\
           &=T[-\cos\phi\ d\theta\ (\delta{x})_z-\sin\phi\ d\theta\ (\delta{y})_z]+dz\left(\frac{t_x}{T}(\delta\ t_x)_z+\frac{t_y}{T}(\delta\ t_y)_z\right)\nonumber\\
           &=\kappa\frac{q}{P}\ {dl}[B_x\ (\delta{x})_z+B_y\ (\delta{y})_z]+dz\left(\frac{t_x}{T}(\delta\ t_x)_z+\frac{t_y}{T}(\delta\ t_y)_z\right)
\end{align}
where the last equality follows from Eq.\eqref{6.5}. Hence, from Eq.\eqref{3.13} and Eq.\eqref{3.15}, 
we can express $\delta[d(q/P)]$ in terms of $\delta(dl)$ as:

\begin{align}\label{3.18}
 \delta\left[d(q/P)\right]&=\left\lbrace \frac{f''(l)}{f'(l)}\ dl + \frac{1}{2}\frac{f'''(l)}{f'(l)}\ dl^2\right\rbrace\delta({\frac{q}{P}})_l\nonumber\\
                          &+\left[f'(l)+f''(l)\ dl\right]\delta(dl)
\end{align}
Hence, the error propagation for $q/P$ may be given as the following:

\begin{align}\label{3.19}
\delta(q/P)_{l+dl}&=\left[1+\left\lbrace \frac{f''(l)}{f'(l)}\ dl + \frac{1}{2}\frac{f'''(l)}{f'(l)}\ dl^2\right\rbrace\right]\delta(q/P)_l\nonumber\\
                  &+ \kappa(f'+f''\ dl)\ f(l)\ T\ dl\left[-B_y\  (\delta{x})_l+B_x\ (\delta{y})_l\right]\nonumber\\
                  &+ (f'+f''\ dl)\ dz\left(\frac{t_x}{T}(\delta\ t_x)_l+\frac{t_y}{T}(\delta\ t_y)_l\right)
\end{align}
We have used central difference formulae of order four (Richardson extrapolation) for 
calculating the derivatives $f', f'', f'''$ etc. with the help of ~\cite{Groom:2001kq}. 
We found that the convergence of $\frac{q}{P}$ to the desired values is very sensitive 
to the calculation of $\frac{\partial(\frac{q}{P})_{l+dl}}{\partial(\frac{q}{P})_{l}}$ 
term.

\subsection{Random Error Contribution}\label{Q}
So far we have discussed the average deterministic error terms. However, as indicated in 
Eq.\eqref{2.4}, we must also account for the random errors due to multiple scattering and 
energy loss straggling. The multiple scattering matrix accounts for the random variation 
of the track elements $(x,y,t_x,t_y)$ over the average motion controlled by the magnetic 
field. 

After crossing a distance $d$ the average angular variance in the track direction due to 
multiple scattering is given by~\cite{fruhwirth2001quantitative}:

\begin{equation}\label{msv}
 \langle\theta_{ms}^2\rangle=\frac{225\times10^{-6}}{\beta^2P^2}\frac{d}{X_s}
\end{equation}
where $X_s$ is related to the radiation length of the material. We used the formula given by Eq.\eqref{msv} that keeps the angular variance proportional to the thickness $d$. The expression 
given by Lynch~\cite{lynch1991approximations} was used in previous work~\cite{MuLkUpTbl} 
to account for multiple scattering. But the presence of logarithmic term in that formula 
makes the calculation dependent on the step size, which is undesirable~\cite{fontana2007track}.
We applied the thick scatterer approximation to iron. The scattering from dense components 
of RPC like glass, copper and aluminium were also considered.

Apart from these, the fluctuations in the rate of energy loss also affect the motion of 
the muon to a great extent. The resulting muon energy loss distribution is a Landau (or 
a Vavilov) distribution. The corresponding random error propagation is done by using the 
covariance terms: $\rm cov[\xi,q/P]$, where $\xi\in(x, y, t_x, t_y,q/P)$. The term $\rm 
cov[q/P,q/P]$ is related to the variance of the truncated Landau distribution through~\cite{fontana2007track}:

\begin{align}
 \rm cov[q/P,q/P]&=\frac{1}{P^4}\sigma^2(P)\nonumber\\
             &=\frac{E^2}{P^6}\sigma^2(E),
\end{align}
We have evaluated $\sigma(E)$ from the Urban model~\cite{lassila1995energy} by sampling 
the number of of collisions suffered by the particle from a Poisson's distribution (parametrized by the mean excitation potential of the material, the fraction $\alpha$ 
corresponding to the area of the truncated Landau distribution, and other parameters of 
the model). The fraction $\alpha$ was found to span rather wide range $(0.993$-$0.999)$ if 
we wish to obtain unit standard deviation of the $q/P$ pull distribution in wide $P_\mu$
-$\theta_\mu$ range. In this paper, we report results obtained by setting $\alpha=0.998$.
The covariance term $\rm cov(x,q/P)$ was derived as below:

\begin{align}
\rm cov[x,q/P]&\equiv\rm cov[q/P,x]\nonumber\\
              &=\frac{\partial{x(P)}}{\partial{P}}\frac{\partial}{\partial{P}}\left[\frac{q}{P}\right]\sigma^2(P)\nonumber\\
              &=-\frac{q}{P^2}\frac{\partial{x(P)}}{\partial{P}}\sigma^2(P),
\end{align}
where $x=x(P)$ is known from Eq.\eqref{3.1}. Other cross terms were also evaluated this 
way using Eq.\eqref{3.2}-\eqref{3.4}. Here we mention that $\langle\theta_{ms}^2\rangle
$ and $\sigma(E)$ are already implemented in the free C++ track fitting library GENFIT~\cite{hoppner2010novel}, 
used by PANDA collaboration. The deterministic propagation of the Kalman parameters in 
GENFIT is based on EMC internal reports~\cite{Wittek:931170}.

\section{Results}\label{results}

In this section, we shall show the performance of the new Kalman filter algorithm coded 
on the basis of the formulae described in section~\ref{propagator}. For this, 5000 muon 
($\mu^\pm$) tracks were simulated through virtual ICAL detector constructed with GEANT4
~\cite{Agostinelli2003250} tool-kit. The following study has been performed for muons of 
generator level momenta $P_\mu\in[1.0-10.0]$ GeV/c at $\cos\theta_\mu=0.95, 0.75, 0.55$.
The vertices of these tracks were uniformly smeared 
over the volume $(43\rm{m}\times14.4\rm{m}\times10\rm m)$ around the centre of the ICAL 
detector. This includes a major region of non-uniform magnetic field (Fig.~\ref{f1(b)}). 
An average strip multiplicity of 1.3 was taken in this work. The uncorrelated noise was
enabled inside the detector. Hit detection efficiency of the central $1.88\rm{m}\times1
.88\rm{m}$ area of each RPC was assumed to be 99\%. Rest portion of the RPC was assumed 
to belong to dead space.

\subsection{Goodness of Fits}%\FloatBarrier
The goodness of typical fits is expressed by the Pull distributions and the chi squares 
($\chi^2$) of the fits. The pull of a fitted parameter $x$ is defined as: 

\begin{equation}\label{5.1}
 P(x)=\frac{x_{reco}-x_{true}}{\sqrt{C_{xx}}}
\end{equation}
Here $x_{true}$ is the true MC value and $x_{reco}$ is the reconstructed value. $C_{xx}$ 
is the diagonal element of the error covariance matrix, corresponding to the element $x$.
\FloatBarrier
\begin{figure}[ht]

\centering
\subfigure[Pull of X]
{
  \includegraphics[width=0.46\textwidth,height=0.315\textwidth]{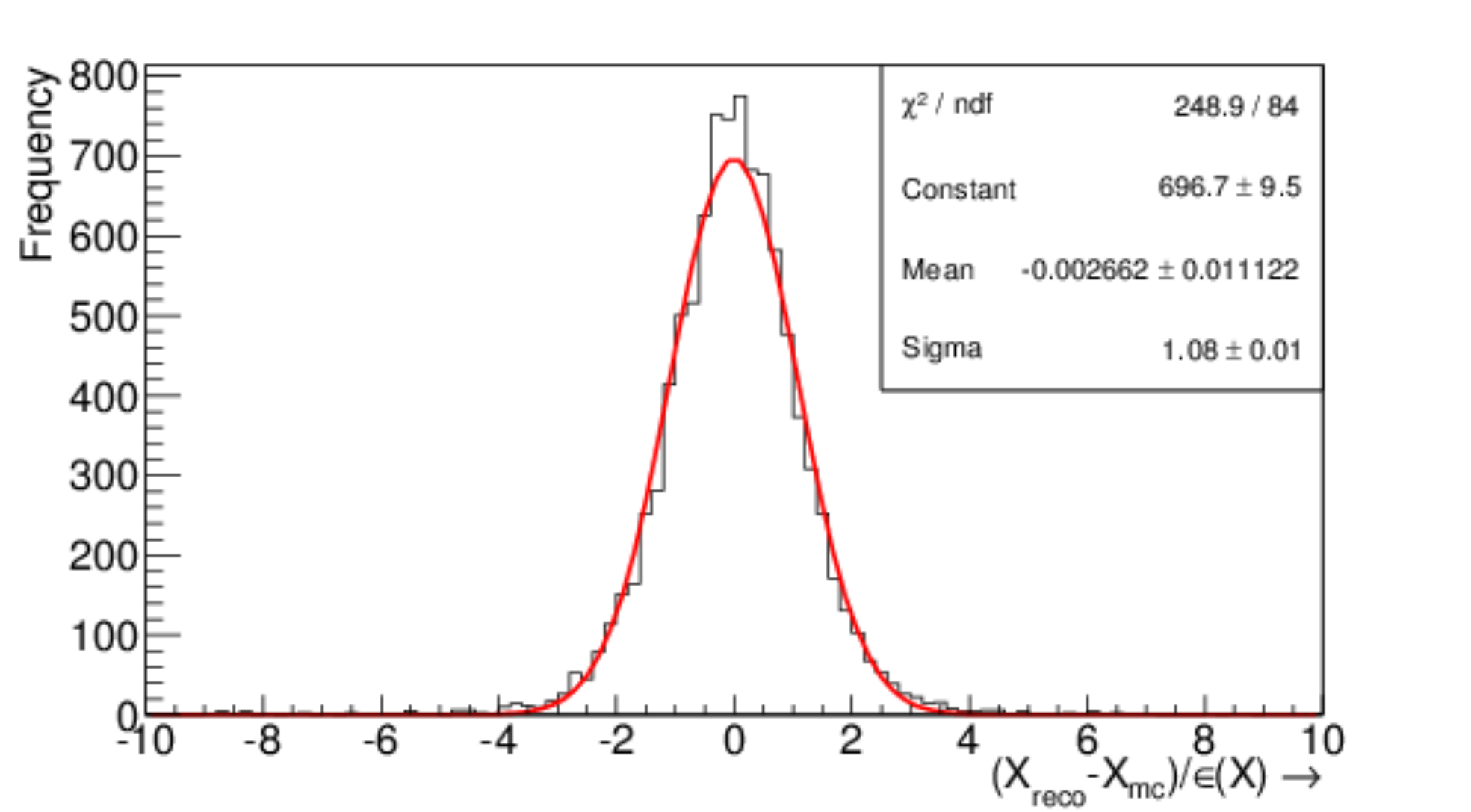}
  \label{Figure:3a}
}
\hspace{0.01 cm}
\subfigure[Pull of Y]
{
  \includegraphics[width=0.46\textwidth,height=0.315\textwidth]{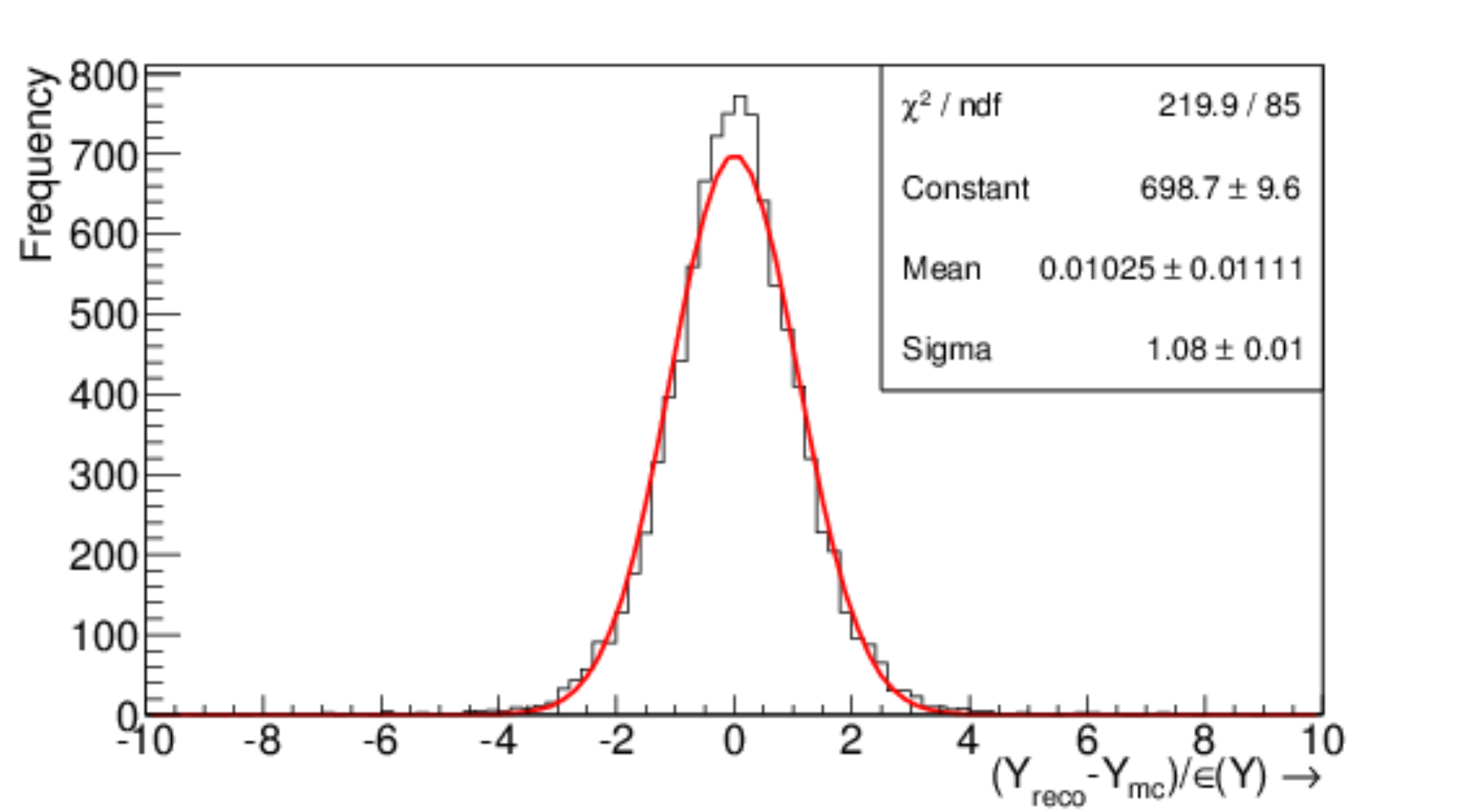}
  \label{Figure:3b}
}
\subfigure[Pull of $t_x$]
{
  \includegraphics[width=0.46\textwidth,height=0.315\textwidth]{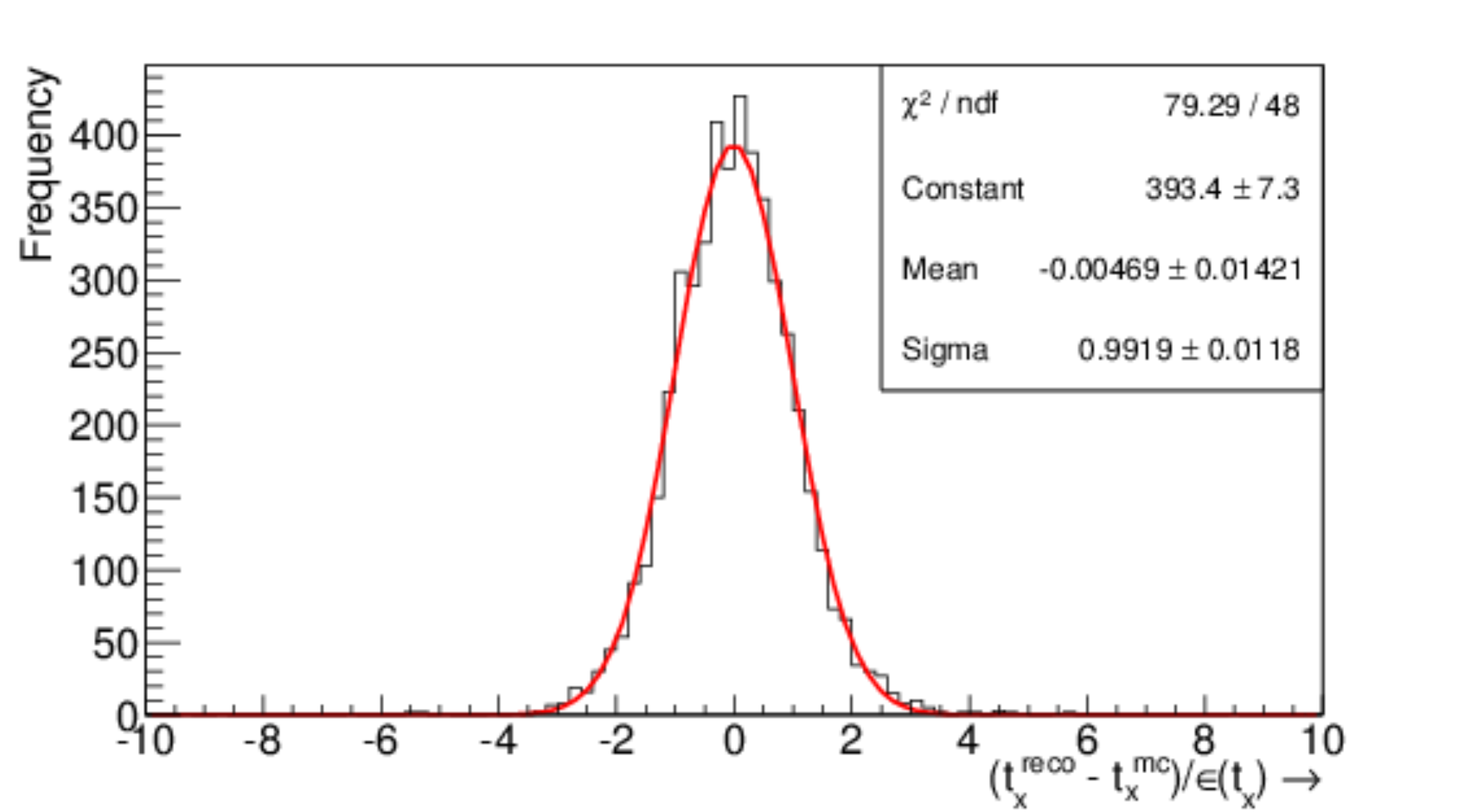}
  \label{Figure:3c}
}
\hspace{0.01 cm}
\subfigure[Pull of $t_y$]
{
  \includegraphics[width=0.46\textwidth,height=0.315\textwidth]{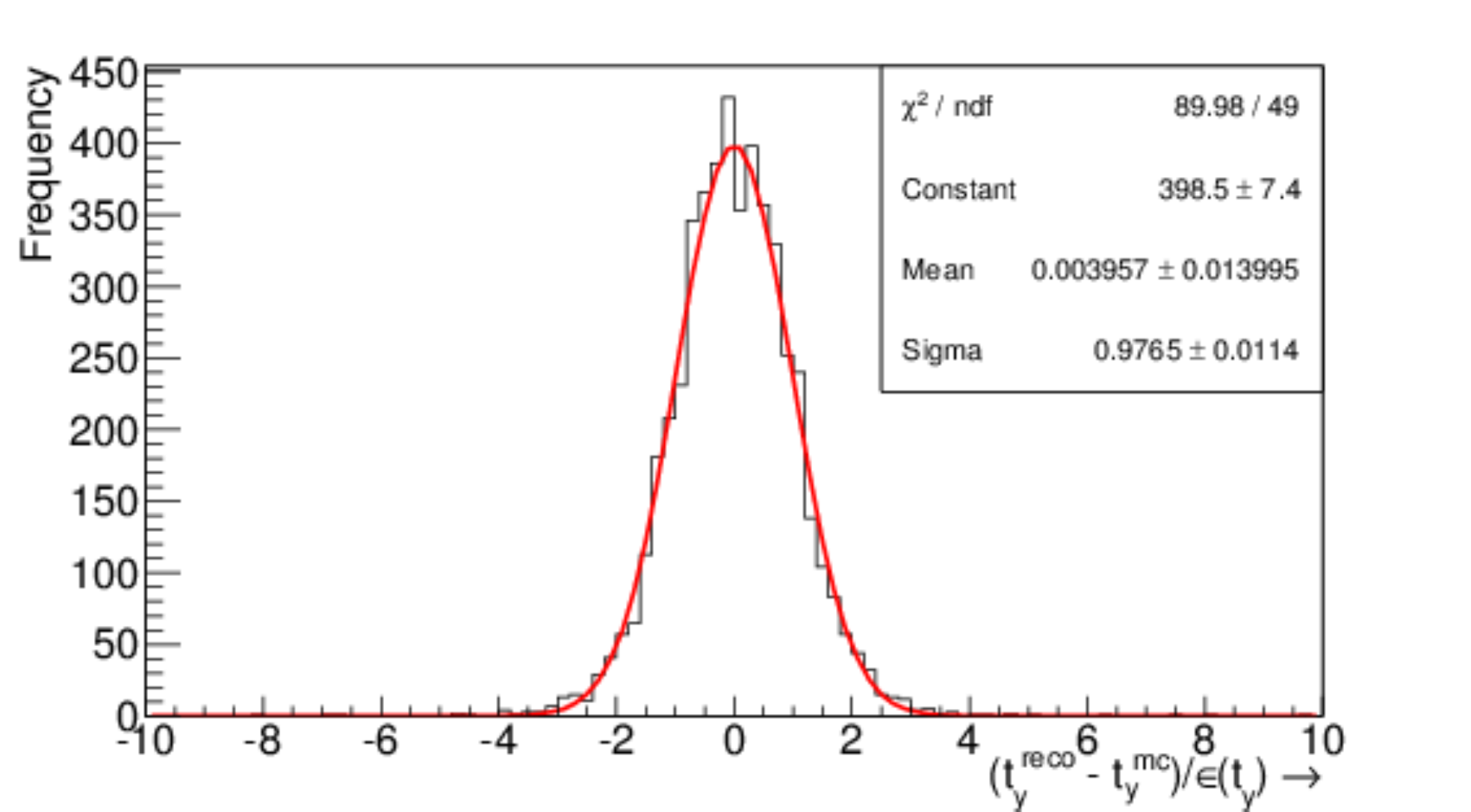}
  \label{Figure:3d}
}
\subfigure[Pull of $q/P$]
{
  \includegraphics[width=0.46\textwidth,height=0.315\textwidth]{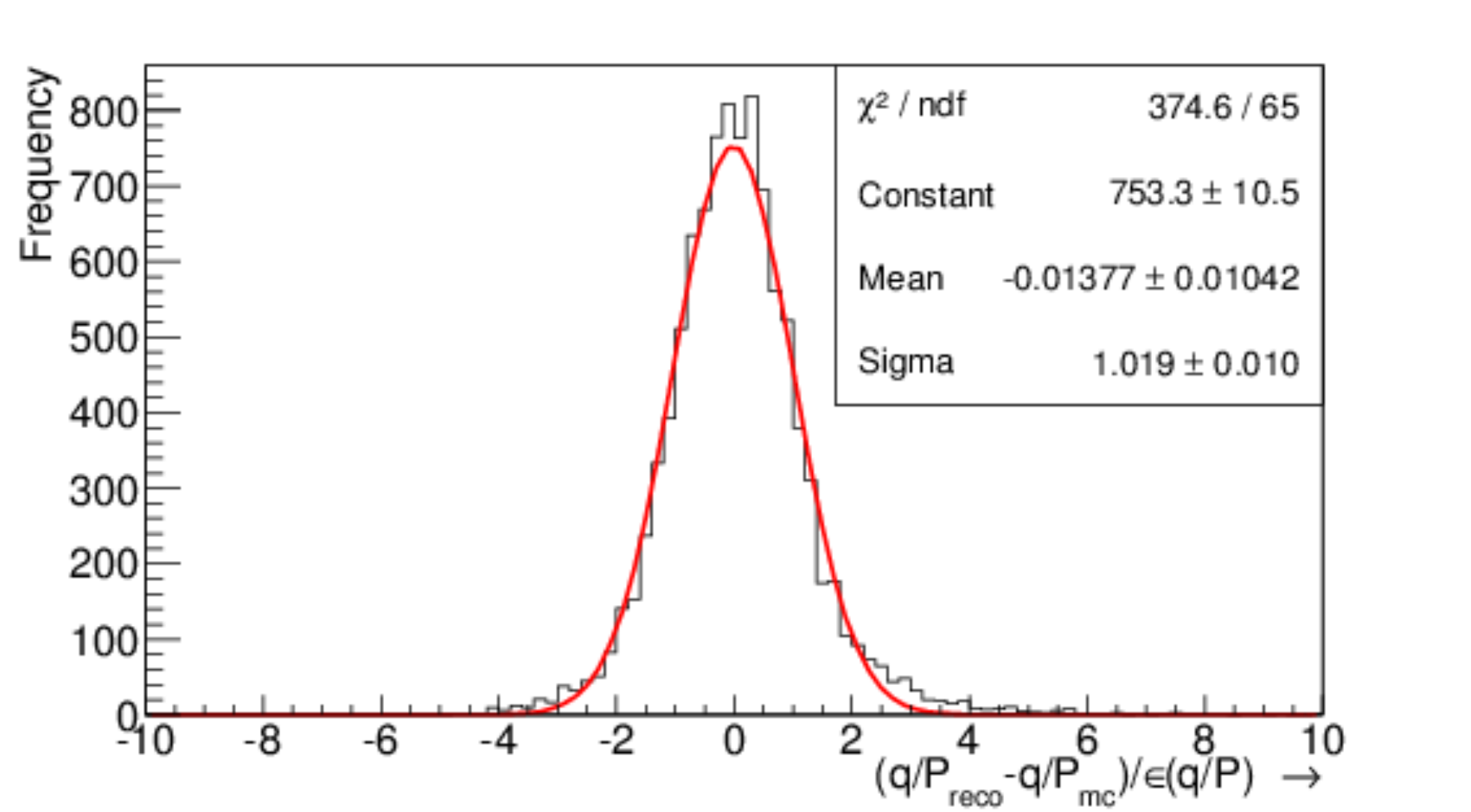}
  \label{Figure:3e}
}
\hspace{0.01 cm}
\subfigure[p-Value Distribution]%Reduced $\chi^2$
{
  \includegraphics[width=0.46\textwidth,height=0.315\textwidth]{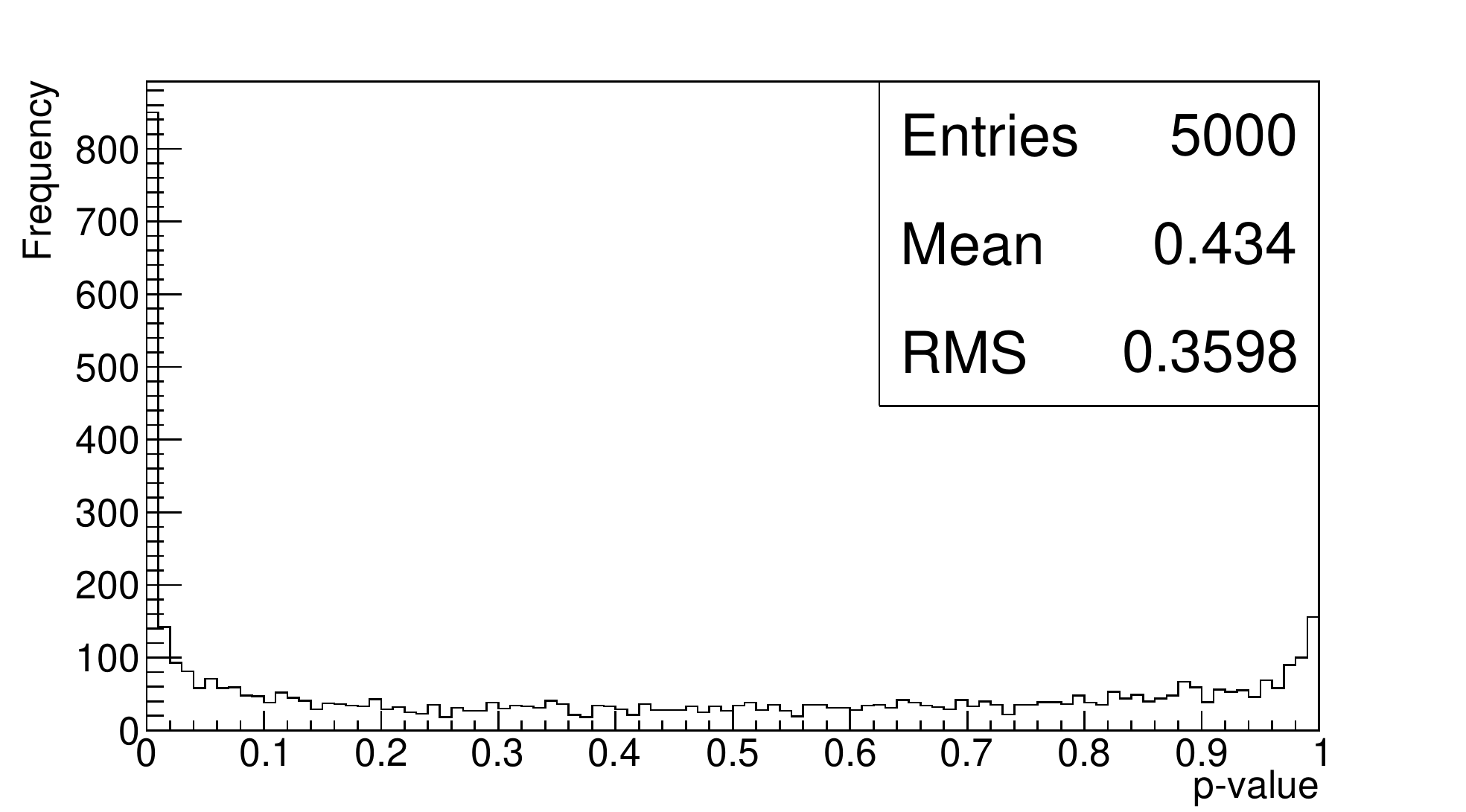}%6GeVchisq.pdf
  \label{Figure:3f}
}
\caption{Reconstructed muon of momentum 6 GeV/c at zenith angle $18.2^o$(cos$\theta$=0.95). 
Pull Distributions: (a) X, (b) Y, (c) $t_x$, (d) $t_y$, (e)$\frac{q}{P}$ and (f) p-Value Distribution}
\label{Figure3}
\end{figure}
\FloatBarrier
For good fit, the pull distributions will have mean at zero and standard deviation equal 
to unity and the reduced $\chi^2$ distributions will have mean equal to unity. Apart 
from that, the p-value distributions of total $\chi^2$ should be reasonably uniform in the 
range [0-1.0]. This happens when the shape of the fitted $\chi^2$ distribution approaches 
a true $\chi^2$ probability distribution function. For this to happen, the prediction 
model should be good enough, the measurement errors should be Gaussian distributed and 
the measurements should be independent of each other.

In Fig.~\ref{Figure3}, we show these distributions for $\mu^{-}$ tracks of momentum $6$ 
GeV/c at zenith angle $\cos\theta=0.95$. The error of an element $x$ has been represented 
by $\epsilon(x)$ in these figures. We see that the means of the pull distributions are 
close to zero and their standard deviations are close to one. The p-value distribution 
is more or less uniform in the range [0.1-0.8]. The peak near zero ($p\rightarrow0^+$) 
comes from the events lying in the tail of the fitted reduced $\chi^2$ distribution. The 
prediction model is not good enough for such tracks. On the other hand, the small heap 
near unity comes from events for which the measurements are highly correlated because of 
multiple scattering. This correlation leads to larger error ($V_k + H_k C_k^{k-1} H^T_k$
) which leads to smaller $\chi^2$ value. Indeed, we have seen that the heap becomes less 
prominent at higher momenta [$\ge10$ GeV/c] tracks for which the multiple scattering is 
less dominant. The reduced $\chi^2$ distributions were seen to have mean close to unity. 

\subsection{Performance at various input momentum ($P_\mu$) and zenith angle ($\cos\theta_\mu$)}
\begin{figure}[ht]
\centering
\subfigure[$\langle{\frac{(q/P)_{rec}-(q/P)_{gen}}{\sqrt{C_{{q/P}{q/P}}}}}\rangle$]
{
  \includegraphics[width=0.45\textwidth,height=0.3\textwidth]{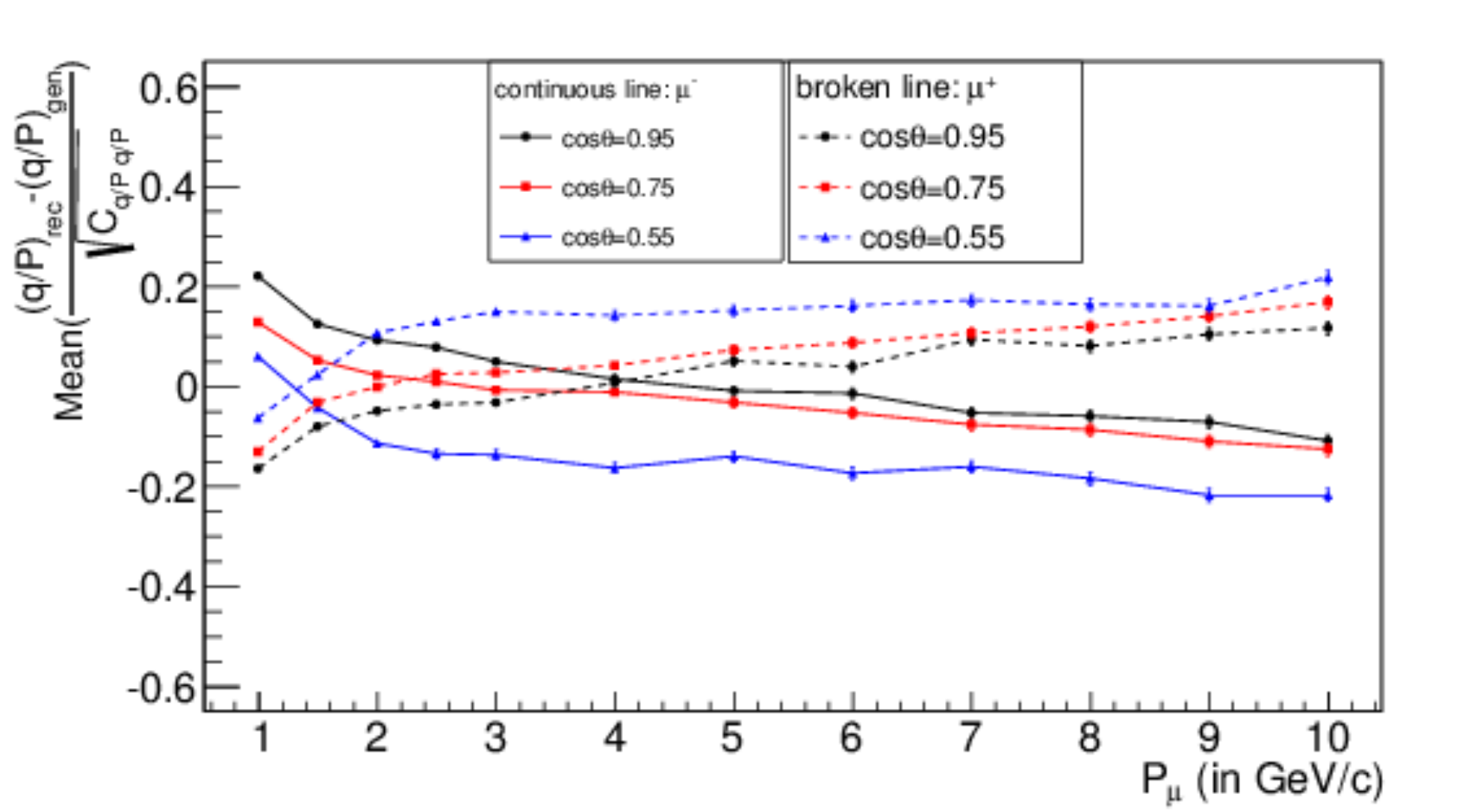}
  \label{f:4(a)}
}
\hspace{0.05 cm}
\subfigure[$\sigma\left({\frac{(q/P)_{rec}-(q/P)_{gen}}{\sqrt{C_{{q/P}{q/P}}}}}\right)$]
{
  \includegraphics[width=0.45\textwidth,height=0.3\textwidth]{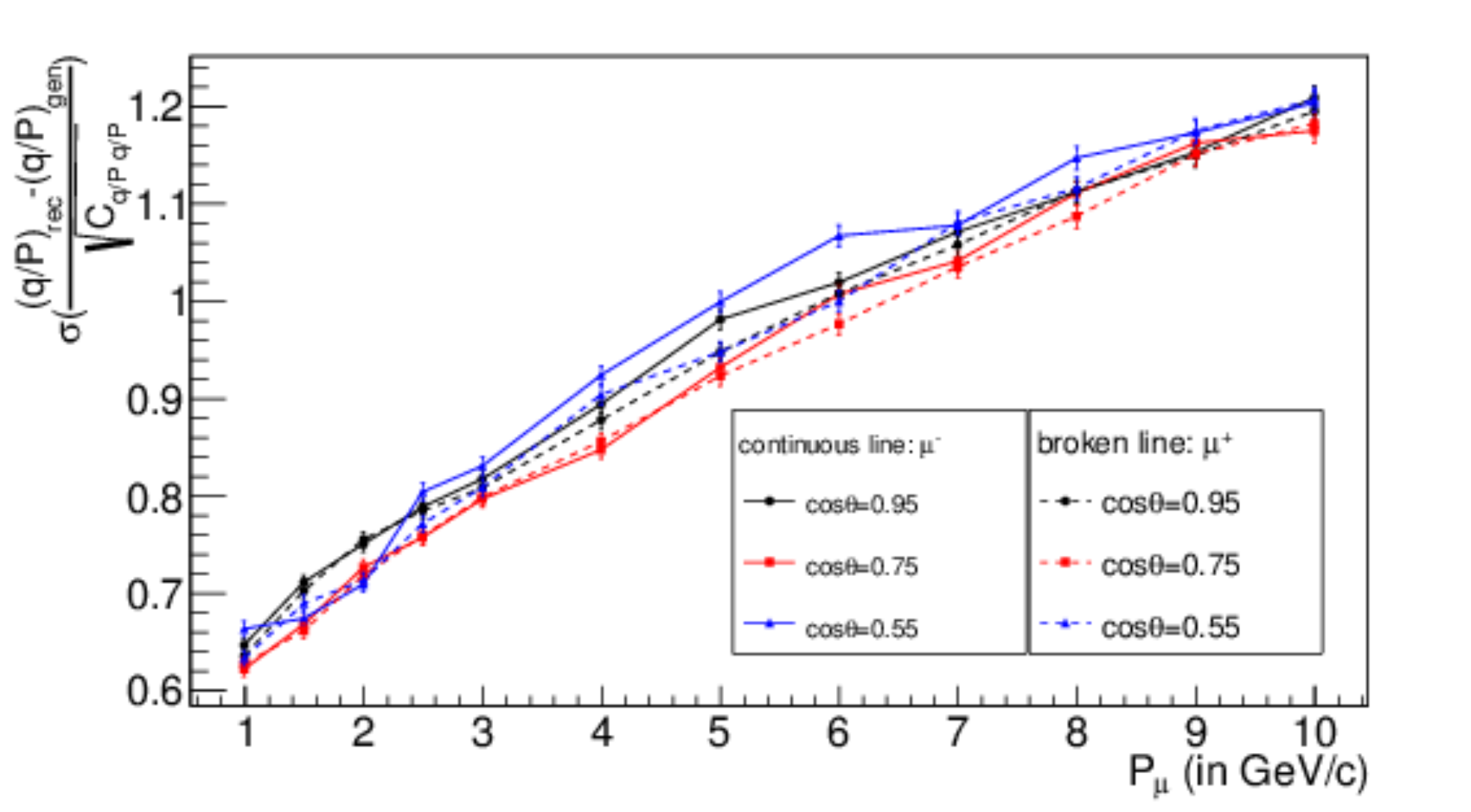}
  \label{f:4(b)}
}
\caption{Variation of Mean and Sigma of q/P Pull Distribution}
\end{figure}

\begin{figure}[ht]
\centering
\subfigure[Charge Identification Efficiency]
{
  \includegraphics[width=0.45\textwidth,height=0.3\textwidth]{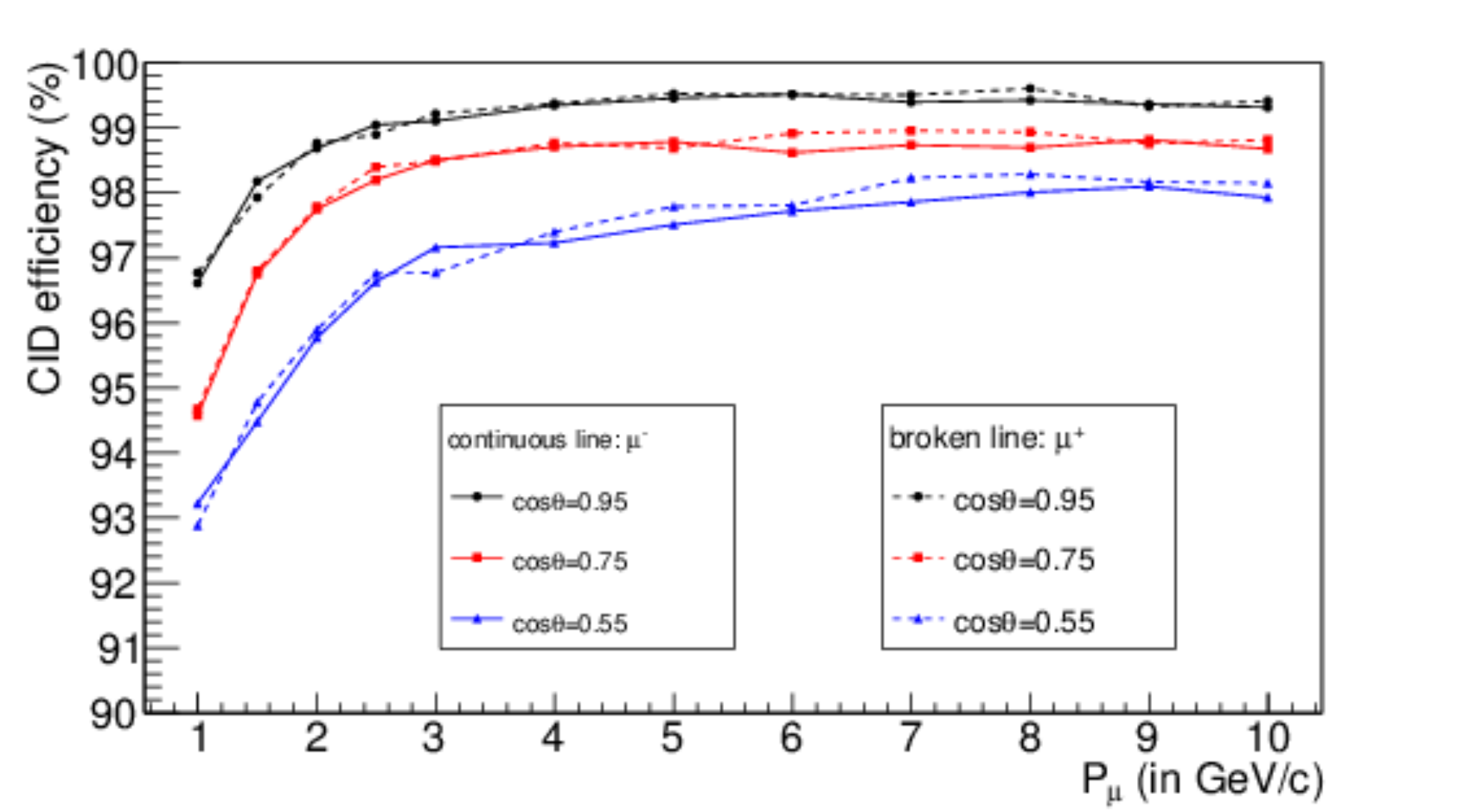}
  \label{f:5(a)}
}
\hspace{0.05 cm}
\subfigure[$\mu^-$ track at $\cos\theta_\mu=0.35$]
{
  \includegraphics[width=0.45\textwidth,height=0.3\textwidth]{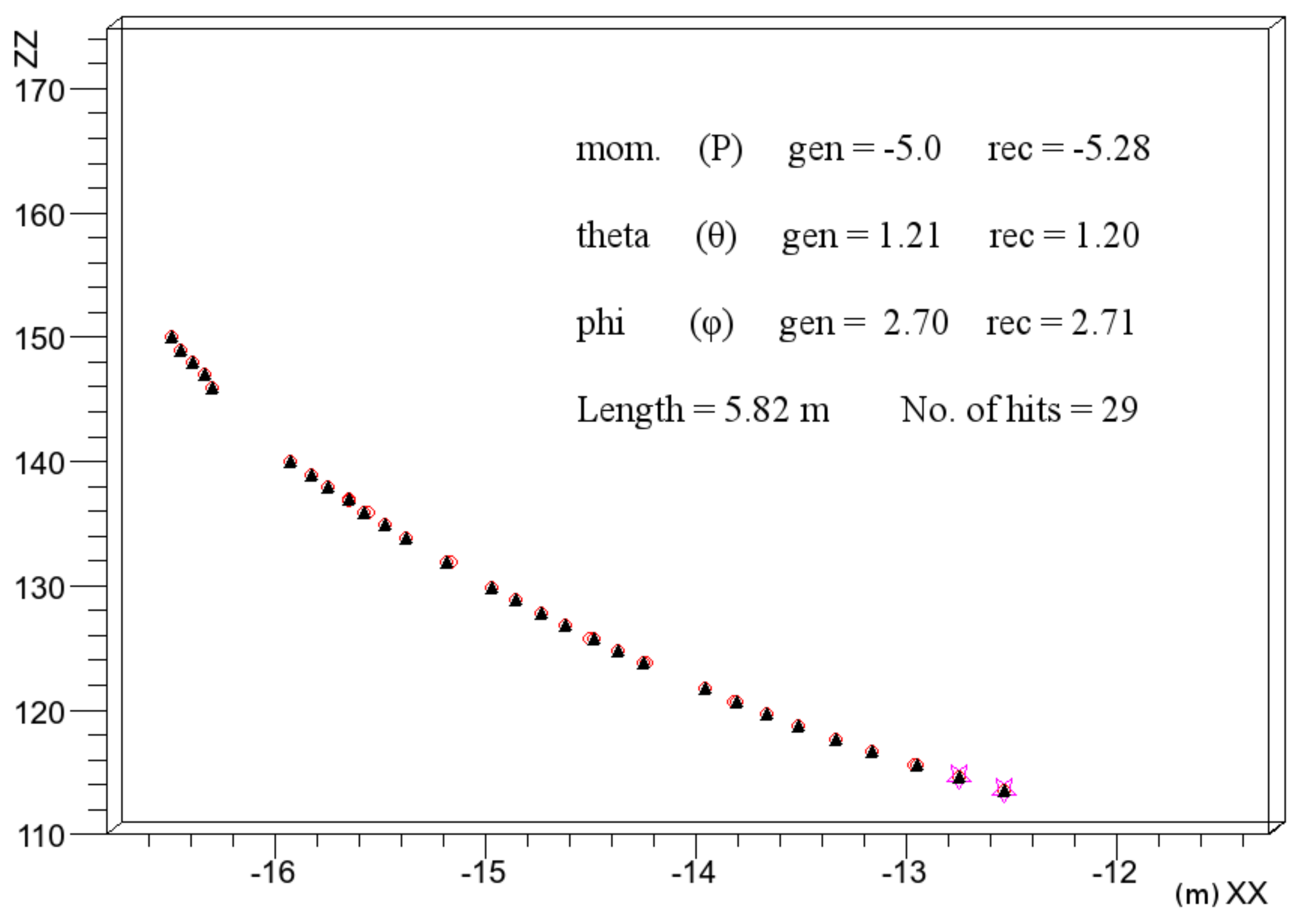}
  \label{f:5(b)}
}
\caption{(a) Variation of Charge Identification Efficiency, (b) A $\mu^-$ 
track at higher zenith angle $\theta_\mu=69.5^o$ with good filter convergence. 
Momentum is in GeV/c and $\theta$ and $\phi$ are in radians. Active RPC planes 
have been plotted along $z$ axis}
\end{figure}

In this subsection, we show the variations of the means and the widths of the $q/P$ Pull 
distributions for various input $P_\mu$ and $\cos\theta_\mu$ values (Fig.~\ref{f:4(a)},
~\ref{f:4(b)}). The corresponding charge identification efficiencies (CID(\%)) is shown 
in Fig.~\ref{f:5(a)}. CID(\%) has been defined as the \% of events (having reduced chi 
square $<2.5$) with correct charge identification.

From Fig.s~\ref{f:4(a)},~\ref{f:4(b)} and ~\ref{f:5(a)}, we see that the Kalman filter convergence of $q/P$ element is more 
or less comparable for $\mu^-$ and $\mu^+$. The mean shift for them is symmetric about 
zero. In fact, as the zenith angle $\theta_\mu$ is increased, the accuracy of the $q/P
$ convergence worsens gradually. The CID efficiency (Fig.~\ref{f:5(a)}) also becomes 
poorer at higher $\theta_\mu$ (lower $\cos\theta$). On the other hand, if the generator 
momenta $P_\mu$ is increased, the standard deviation of the $q/P$ pull distribution 
shows almost linear (monotonic) increase. As mentioned before, Urban model~\cite{lassila1995energy} parameter $
\alpha$ (section~\ref{Q}) was chosen such that the standard deviation of $q/P$ pull distribution 
comes close to unity for $P_\mu\in[4,9]$ GeV/c (as neutrino mass hierarchy sensitivity 
mostly comes from the muons of such momenta). The Means and standard deviations of the 
pull distributions for $(x, y, t_x, t_y)$ were also found to be consistent over a wide 
range of $P_\mu$ and $\cos\theta_\mu$. Fig.~\ref{f:5(b)} shows a $\mu^-$ track of $P
_\mu=5$ GeV/c at $\cos\theta_\mu=0.35$ (that is, $\theta_\mu=69.5^o$, measured from the 
perpendicular to the horizontal RPC planes). The filter convergence is found better if 
good enough number of measurements can be done along a track.

\subsection{Comparison with the previous code}
Here we shall compare the performance of the new algorithm with that of the existing 
code~\cite{MuLkUpTbl} and see what we have gained from the new algorithm. The following 
Fig.~\ref{Fig:6(a)} shows the distribution of the reconstructed momenta obtained from 
the two codes for input momentum $P_\mu=5$ GeV/c at $\cos\theta=0.95$. Fig.~\ref{Fig:6(b)} 
shows the corresponding $\cos\theta$ distribution. In the following Table~\ref{Table2}, 
we compare the rms errors of fitted $\theta_\mu$ and $P_\mu$ and charge identification 
efficiency(\%) for input $\cos\theta=0.75$ as obtained from the two codes.

\begin{figure}[ht]
\centering
\subfigure[Comparison of $P_{rec}$]
{
  \includegraphics[width=0.45\textwidth,height=0.3\textwidth]{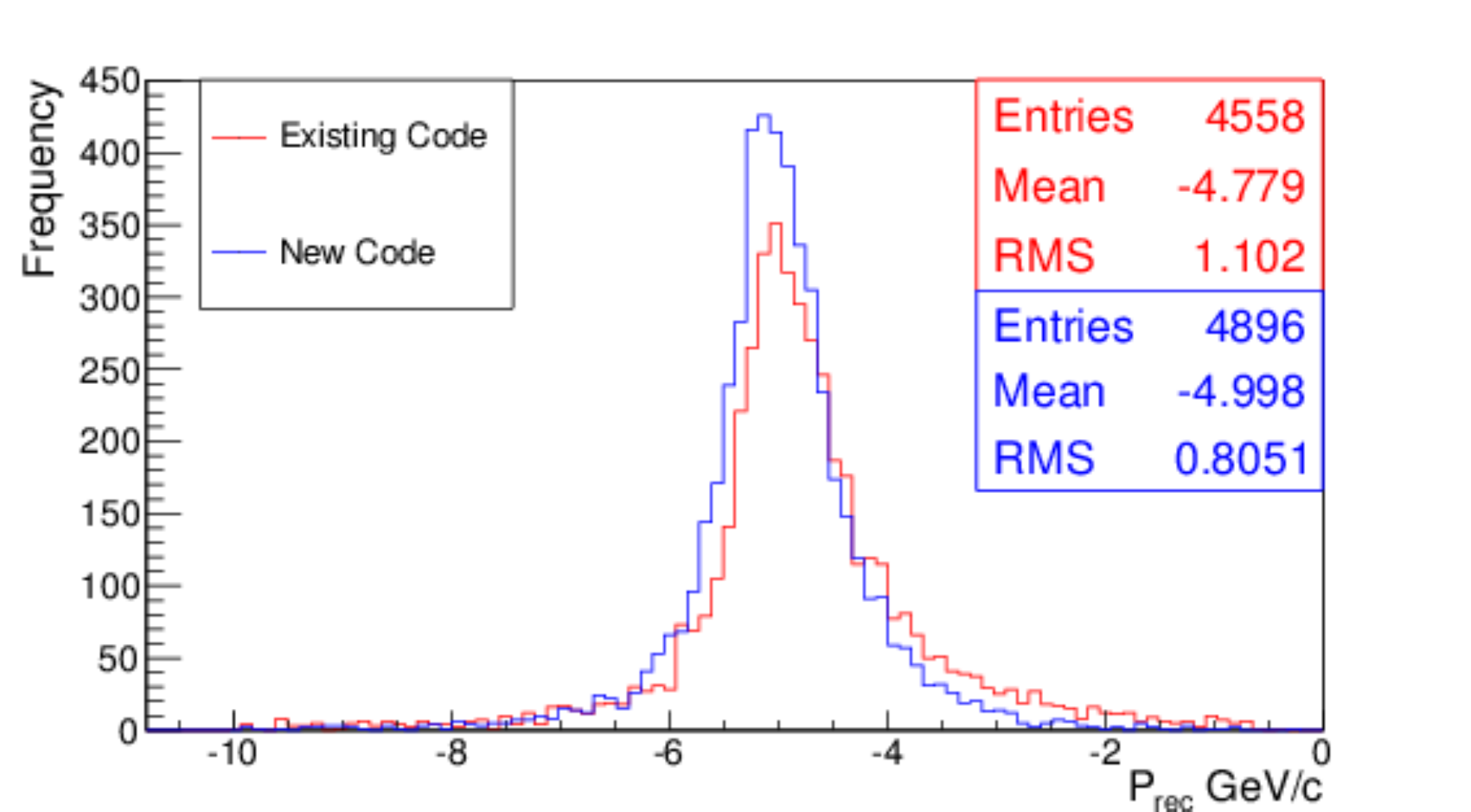}%width=0.45\textwidth,height=0.275\textwidth
  \label{Fig:6(a)}
}
\hspace{0.025 cm}
\subfigure[Comparison of $\cos\theta_{rec}$]
{
  \includegraphics[width=0.45\textwidth,height=0.3\textwidth]{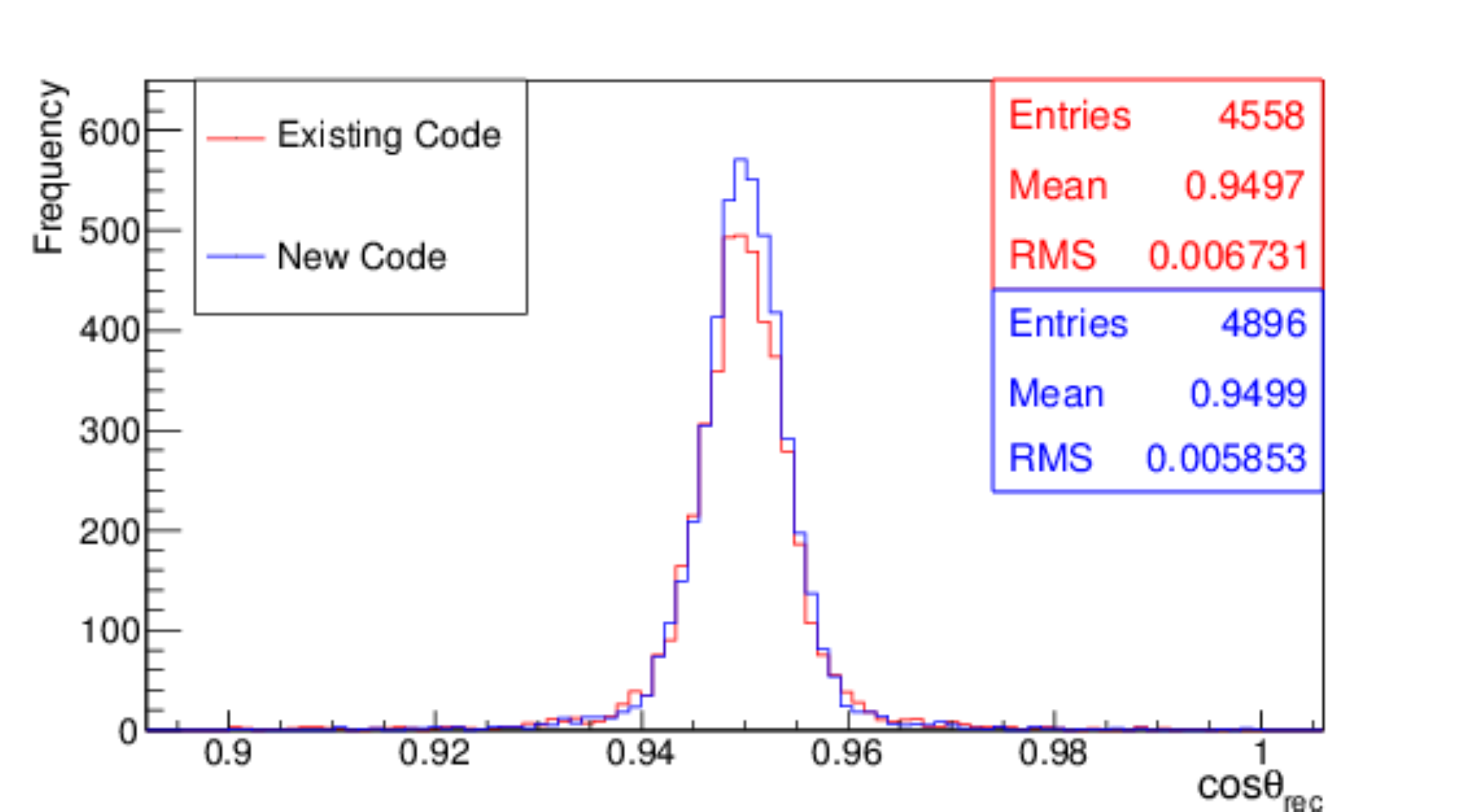}%width=0.45\textwidth,height=0.275\textwidth
  \label{Fig:6(b)}
}
\caption{Comparison of muon reconstruction between Existing Code and New Code}
\end{figure}

\begin{table}[ht]
\caption{Comparison of the two codes:\label{Table2}}                     % title of Table
\centering                         % used for centering table
\begin{tabular}{|c|c||c|c|c||c|c|c||}
\hline
\multicolumn{2}{ |c|| }{Code:-} & \multicolumn{3}{ c|| }{Existing Code} & \multicolumn{3}{ c ||}{Modified Code}\\\cline{1-8}
\hline
P(GeV/c) & cos$\theta$ & $rms(\theta_\mu)$ & $rms(P_\mu)$ & CID\% & $rms(\theta_\mu)$ & $rms(P_\mu)$ & CID\%\\\hline
1.0 & 0.75 & 0.051 & 0.27 & 91.61 & 0.049 & 0.29 & 94.57 \\\hline
2.0 & 0.75 & 0.035 & 0.47 & 96.06 & 0.032 & 0.44 & 97.74 \\\hline
3.0 & 0.75 & 0.029 & 0.61 & 96.07 & 0.026 & 0.57 & 98.50 \\\hline
4.0 & 0.75 & 0.023 & 0.83 & 96.58 & 0.021 & 0.72 & 98.70 \\\hline
5.0 & 0.75 & 0.019 & 0.95 & 96.61 & 0.018 & 0.85 & 98.78 \\\hline
6.0 & 0.75 & 0.016 & 1.11 & 96.57 & 0.016 & 1.03 & 98.61 \\\hline
7.0 & 0.75 & 0.014 & 1.34 & 97.53 & 0.015 & 1.23 & 98.73 \\\hline
8.0 & 0.75 & 0.015 & 1.71 & 97.27 & 0.013 & 1.46 & 98.69 \\\hline
9.0 & 0.75 & 0.015 & 1.86 & 96.93 & 0.012 & 1.69 & 98.81 \\\hline
10.0 & 0.75 & 0.010 & 2.21 & 97.47 & 0.010 & 1.81 & 98.67 \\\hline
\end{tabular}
\label{table:b}                    % is used to refer this table in the text
\end{table}

From the Fig.s~\ref{Fig:6(a)},~\ref{Fig:6(b)} and the Table~\ref{Table2}, 
it is seen that implementing the algorithm (section~\ref{propagator}) gave 
us considerably better results in terms of charge identification efficiency, 
accuracy and precision of the MC input parameters, compared to the previous 
code. The efficiency of reconstruction also increased.

\subsection{Observations:}
\subsubsection{Fully/Partially Contained events}
\begin{figure}[ht]
\centering
\subfigure[]
{
  \includegraphics[width=0.45\textwidth,height=0.3\textwidth]{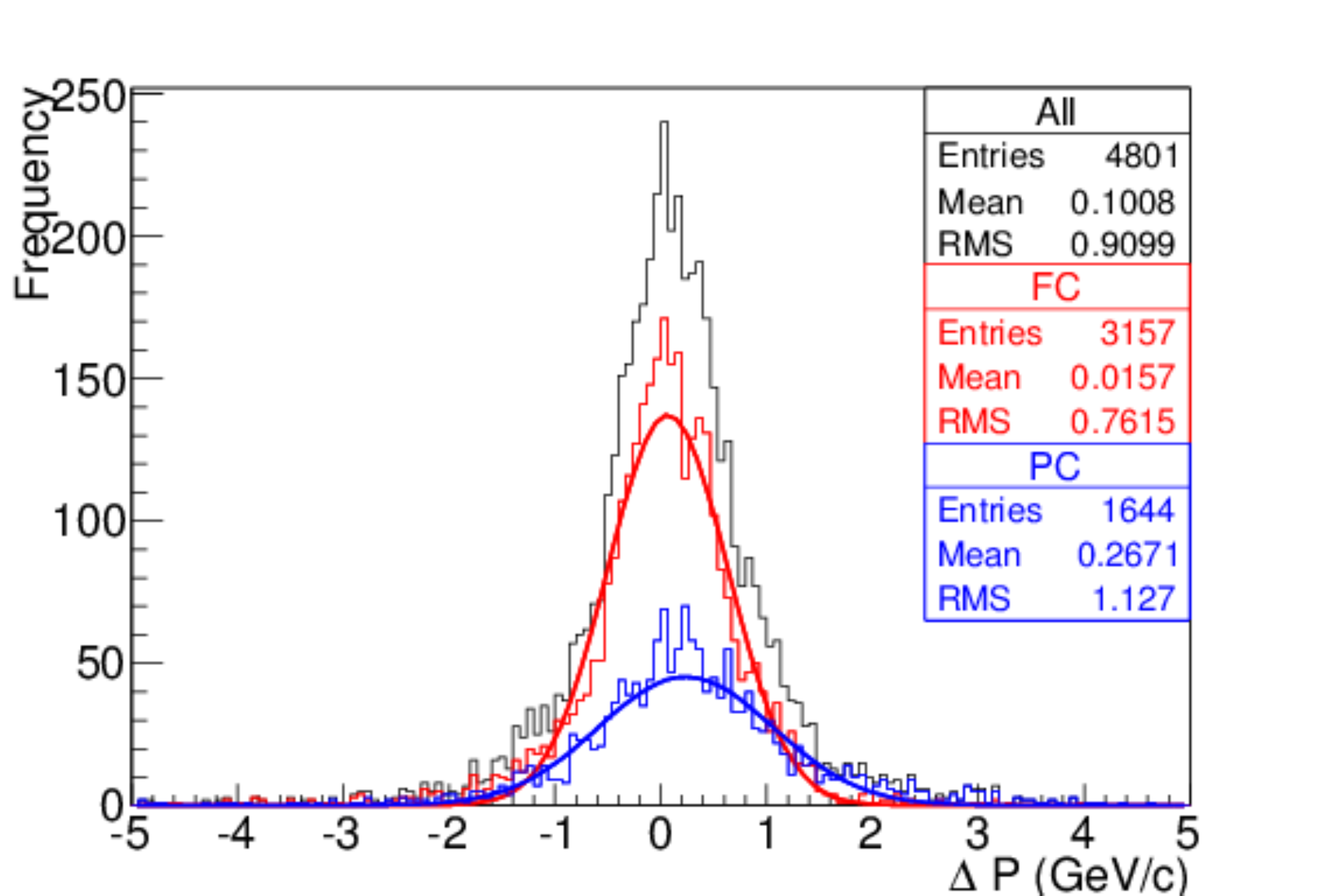}
  \label{Fig:7(a)}
}
\hspace{0.025 cm}
\subfigure[]
{
  \includegraphics[width=0.45\textwidth,height=0.3\textwidth]{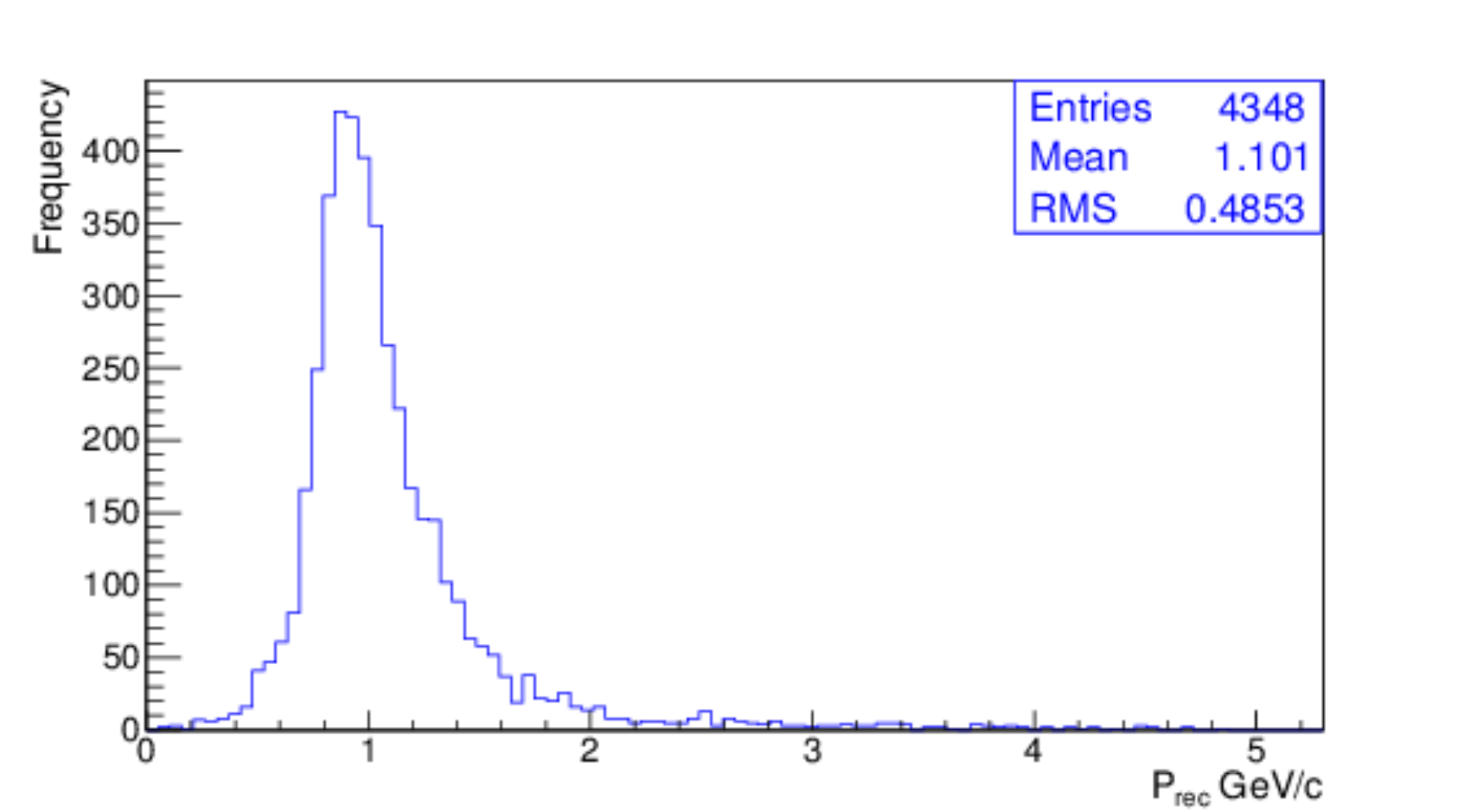}
  \label{Fig:7(b)}
}
\caption{Track Fitting Performance for (a) FC/PC $\mu^-$ Tracks at $P_\mu=5$ GeV/c, 
$\cos\theta_\mu=0.55$; (b) $\mu^+$ Tracks at $P_\mu=1$ GeV/c, $\cos\theta_\mu=0.95$}
\end{figure}

The reconstructed momentum distribution shows tails in both sides, more prominent 
at higher input momenta. They were found mostly due to the presence of Partially 
Contained (PC) tracks - for them, the resolution is poorer. In Fig.~\ref{Fig:7(a)}, 
we show the distribution for Fully Contained (FC) tracks superimposed to the 
Partially Contained (PC) tracks, for muons with momentum $P=5$ GeV/c at $\cos\theta_\mu=0.55$. As 
expected, rms is much better for FC tracks. In Fig.~\ref{Fig:7(a)}, $\Delta{P}$ 
denotes $(P_{reconstructed}-P_{generator})$.

\subsubsection{Behaviour at low Momentum}
The convergence for low momenta $P_\mu<1.5$ GeV/c events are usually poor, because 
in these cases less than $10-12$ hits are available. In Fig.~\ref{Fig:7(b)}, the 
reconstructed momentum distribution of $\mu^+$ of $P_\mu=1.0$ GeV/c at $\cos\theta
_\mu=0.95$ is shown. Non-Gaussian tails are observed at such low momenta leading to 
poor momentum resolution.

\section{Summary}
In this work, we have tried to improve the existing muon track following code for 
ICAL experiment. That algorithm was based on a simple propagator matrix that had 
only first order correction terms~\cite{marshall2008study}. We implemented higher 
order terms in random and deterministic error propagation formulae as obtained in 
section~\ref{propagator}. The code is seen to produce better results compared to 
the existing code. The tail in the momentum distribution is suppressed and hence, 
the momentum resolution has improved. This is visible from Fig.~\ref{Fig:6(a)}. 
The reconstruction efficiency and charge identification efficiency also increased.

We have also performed the consistency checks on the final covariance matrix (that 
the matrix should be symmetric, the diagonal terms should be positive etc.). These 
checks turned out to be fine. It is seen that the convergence of Kalman parameters 
is very sensitive to accurate calculation of the fifth column of the propagator $F
_{k-1}$. The filter is seen to work over a wide 
range of momentum $P_\mu$ and direction $\cos\theta_\mu$ of muon. Except for small 
generator momenta ($<1.5$ GeV/c) and/or low $|\cos\theta_\mu|$ ($<0.4$), the filter is 
seen to produce reasonably accurate estimates. The fact that the mean of the $q/P$ 
pull distributions are close to zero in most of the cases implies that the average 
energy loss of the particle in different constituent materials has been accounted 
for correctly. Higher number of measurements (hits) along a track is found to lead 
to better convergence of track parameters.

However, at lower generator momenta and/or lower $|\cos\theta_\mu|$, the number of 
hits are intrinsically small. Apart from that, multiple Coulomb scattering affects 
the track. In such scenarios, the filter performance is worse. Momentum resolution 
becomes poorer and charge identification efficiency degrades. 

\section{Acknowledgment}
We thank Tata Institute of Fundamental Research where the Research and Developmental 
studies for the future ICAL detector is going on. The visualization software designed 
by D. Samuel was very useful. We also thank Y. P. Viyogi and D. Indumathi and S. Chattopadhyay 
for useful suggestions for improving the article. K.B. expresses gratitude to W. Wittek, 
for providing the details of error propagation along a particle trajectory in a magnetic 
field, which was the basis of the computation shown in section~\ref{qperrorprop}. The 
editor and the referee also helped us for improving the quality of the article with 
proper suggestions.

\section{Appendix}
\begin{figure}[h]
\begin{center}
\scalebox{0.5}{\includegraphics{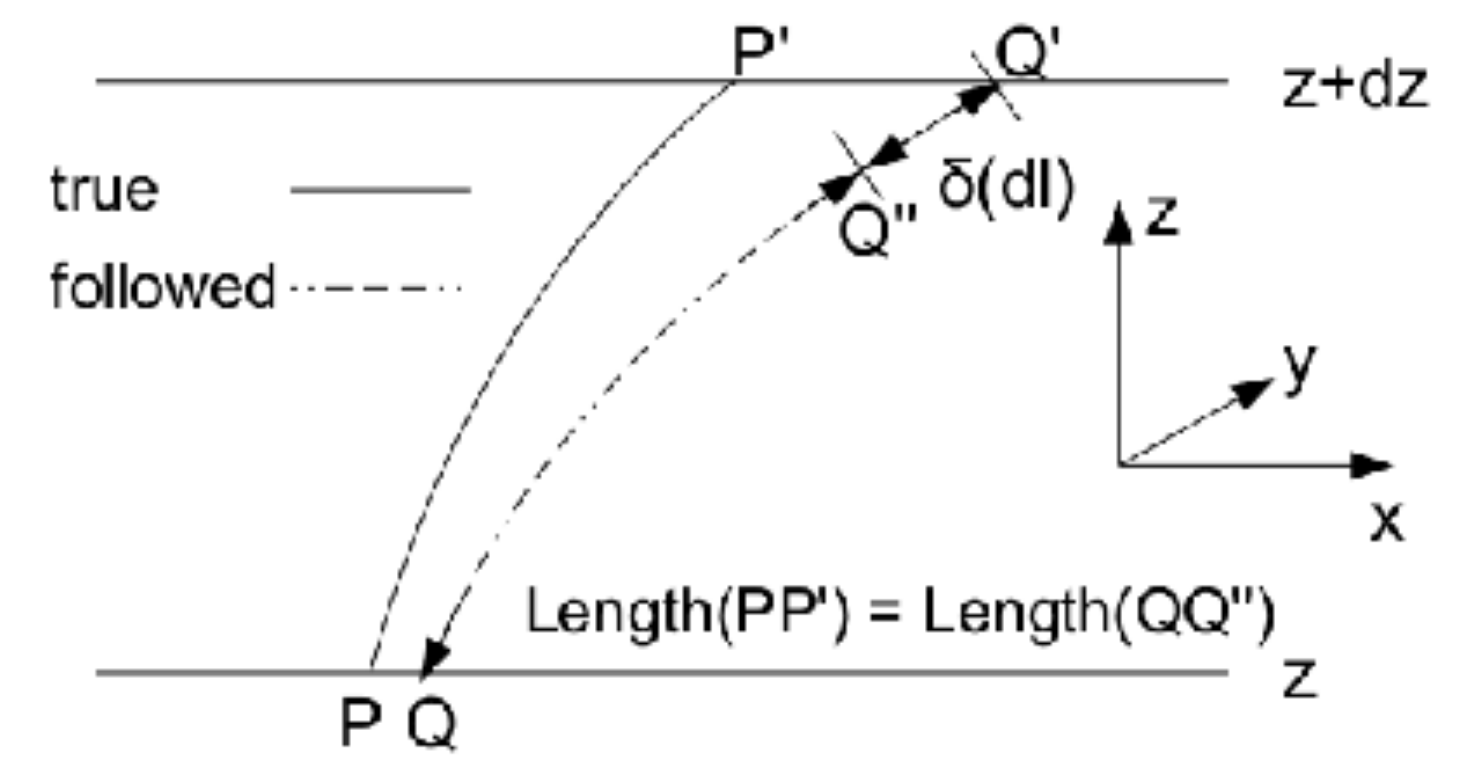}}
\caption{\label{fig:10} $\delta(dl)$ Correction}
\end{center}
\end{figure}

The calculation of the propagator matrix in~\cite{Wittek:931170} was facilitated by the 
use of $SC$ coordinate system $(x_\perp,y_\perp,z_\perp)$, where $x_\perp$ is along the 
track direction and $y_\perp$ and $z_\perp$ are chosen to be locally orthogonal to the 
track. In this frame, the detector planes are the planes of constant $x_\perp$, and therefore, 
$\delta(dl)$ in section~\ref{qperrorprop} is exactly equal to $\delta(dx_\perp)$ and 
both $\delta(dy_\perp)$ and $\delta(dz_\perp)$ are equal to zero. This frame is related 
to the Cartesian coordinates by the following equation~\cite{Wittek:931170}:

\begin{equation}\label{6.1}
\begin{bmatrix}
        {x_\perp} \\
        {y_\perp} \\
        {z_\perp}
\end{bmatrix}
=
\begin{bmatrix}
        \cos\lambda\ \cos\phi & \cos\lambda\ \sin\phi & \sin\lambda \\
       -\sin\phi & \cos\phi & 0 \\
       -\sin\lambda\ \cos\phi &-\sin\lambda\ \sin\phi & \cos\lambda
\end{bmatrix}
\begin{bmatrix}
        {x} \\
        {y} \\
        {z}
\end{bmatrix}
\end{equation}
where $\phi$ is the azimuthal angle and $\lambda$ is the dip angle and is related to the zenith 
angle $\theta$ as $\lambda=90^o-\theta$. It has been shown that in the $SC$ system, that the 
infinitesimal displacement of the track in terms of infinitesimal deflections ($d\lambda$, $d\phi$) 
in the magnetic field can be written as:

\begin{equation}\label{6.2}
\begin{bmatrix}
        {x'_\perp} \\
        {y'_\perp} \\
        {z'_\perp}
\end{bmatrix}
=
\begin{bmatrix}
        1 & (\cos\lambda\ d\phi) & d\lambda \\
       -\cos\lambda\ d\phi & 1 & \tan\lambda\ (\cos\lambda\ d\phi)\\
       -d\lambda &-\tan\lambda\ (\cos\lambda\ d\phi) & 1
\end{bmatrix}
\begin{bmatrix}
        {x_\perp} \\
        {y_\perp} \\
        {z_\perp}
\end{bmatrix}
\end{equation}
which corresponds to a deflection of angle ${d\vec{\alpha}}$ (due to magnetic field), given by:%=-\kappa{\bf{B}}\frac{q}{P}dl

\begin{equation}\label{6.3}
d\vec{\alpha}=
\begin{bmatrix}
        {\sin\lambda\ d\phi} \\
        {-d\lambda} \\
        {\cos\lambda\ d\phi}
\end{bmatrix}
% =-\kappa\frac{q}{P}
% \begin{bmatrix}
%         {{B_1}} \\
%         {{B_2}} \\
%         {{B_3}}
% \end{bmatrix}dl
\end{equation}
% where $(B_1,B_2,B_3)$ is the magnetic field in the SC system. 
We want to find the corresponding equations in ICAL. With the help of $3\times3$ Jacobian matrix in Eq.\eqref{6.1}, we do 
similarity transformation of Eq.\eqref{6.2} and obtain:

\begin{equation}\label{6.4}
\begin{bmatrix}
        {x'} \\
        {y'} \\
        {z'}
\end{bmatrix}
=
\begin{bmatrix}
        1 & d\phi & -\cos\phi\ d\theta \\
       -d\phi & 1 & -\sin\phi\ d\theta \\
        \cos\phi\ d\theta & \sin\phi\ d\theta & 1
\end{bmatrix}
\begin{bmatrix}
        {x} \\
        {y} \\
        {z}
\end{bmatrix}
\end{equation}
corresponding to an angle $d{\vec{e}}=\kappa\frac{q}{P}({\bf{e}}\times{\bf{B}})dl
$ [Eq.(8) in~\cite{Gorbunov:2006pe} with ${\bf{e}}=(\frac{t_x}{T},\frac{t_y}{T},
\frac{1}{T})$ where:

\begin{equation}\label{6.5}
d\vec{e}=
\begin{bmatrix}
        {\sin\phi\ d\theta} \\
        {-\cos\phi\ d\theta} \\
        {-d\phi}
\end{bmatrix}
=\kappa\frac{q}{P}\frac{1}{T}
\begin{bmatrix}
        {-{B_y}} \\
        {+{B_x}} \\
        {(t_xB_y-t_yB_x)}
\end{bmatrix}dl
\end{equation}
Eq.\eqref{6.5} says that the direction of the particle of momentum $P$ is rotated by the 
magnetic field $\bf B$ through an angle $d\vec{e}$ over a track length $dl$. As the track 
of the particle is followed from ${{\bf{r}}(z)}$ to ${{\bf{r}}(z+dz)}$ an error $\delta$ in the estimation of the 
differential increment in the particle track length $dl$ happens due to curvature of the 
track in magnetic field (Fig.~\ref{fig:10}). From Eq.\eqref{3.17}, we see that $\delta
(dl)$ depends on $\delta(dz)$.

\begin{equation}\label{6.6}
\begin{bmatrix}
        \delta{x} \\
        \delta{y} \\
        \delta{z}
\end{bmatrix}_{{\bf{r}}(z+dz)}
=
\begin{bmatrix}
        1 & d\phi & -\cos\phi\ d\theta \\
       -d\phi & 1 & -\sin\phi\ d\theta \\
        \cos\phi\ d\theta & \sin\phi\ d\theta & 1
\end{bmatrix}
\begin{bmatrix}
        \delta{x} \\
        \delta{y} \\
        \delta{z}
\end{bmatrix}_{{\bf{r}}(z)}
+
\begin{bmatrix}
        \delta(dx) \\
        \delta(dy) \\
        \delta(dz)
\end{bmatrix}\,
%\label{eq:symmetrical}
\end{equation}
The errors $\delta x$, $\delta y$ and $\delta z$ at ${{\bf{r}}(z)}$ propagate to ${{\bf{r}
}(z+dz)}$ according to Eq.\eqref{6.4}. Apart from these, the errors $\delta(dx)$, $\delta(
dy)$ and $\delta(dz)$ also creep in due to the curvature of the track. The total errors $
\delta x$, $\delta y$ and $\delta z$ at ${{\bf{r}}(z+dz)}$ are concisely given by Eq.\eqref{6.6}.

We found the key relation Eq.\eqref{6.5} from Eq.\eqref{6.2} (valid in SC system) by using 
$(3\times3)$ Jacobian (Eq.\eqref{6.1}) for coordinate transformation between the SC system 
and the Cartesian system. The propagator in the SC system~\cite{Wittek:931172}, derived from 
helix model, was not directly used to express $\delta(dl)$ in terms of $\delta{x}$ etc. In 
fact, the propagator matrix $F_{k-1}$ in section~\ref{propagator} is based on the analytic 
formulae for track extrapolation. Of course, one can obtain the transformed propagator (equipped with 
these formulae) in SC/perigee system by using the $(5\times 5)$ Jacobian matrices (Eq. A24, 
A25, A28, A29) derived in~\cite{strandlie2006derivation}. They are very helpful when $F_{k
-1}$ is not known in the required system, but is known in some other system.

%\nolinenumbers
\bibliographystyle{unsrt}%plain
\bibliography{KFpaper1}

\begin{thebibliography}{10}

\bibitem{1960Kalman}
Rudolph~Emil Kalman.
\newblock A new approach to linear filtering and prediction problems.
\newblock {\em Transactions of the ASME--Journal of Basic Engineering},
  82(Series D):35--45, 1960.

\bibitem{9780521635486}
R.~Fr{\"u}hwirth, M.~Regler, R.~K. Bock, H.~Grote, and D.~Notz.
\newblock {\em Data Analysis Techniques for High-Energy Physics (Cambridge
  Monographs on Particle Physics, Nuclear Physics and Cosmology)}.
\newblock Cambridge University Press, 2000.

\bibitem{fujiiextended}
Fujii Keisuke.
\newblock Extended kalman filter.
\newblock {\em The ACFA-Sim-J Group,}, pages pp. 7, 15--17, 20, 2002.

\bibitem{wan2000unscented}
Eric~A Wan and Rudolph Van Der~Merwe.
\newblock The unscented kalman filter for nonlinear estimation.
\newblock In {\em Adaptive Systems for Signal Processing, Communications, and
  Control Symposium 2000. AS-SPCC. The IEEE 2000}, pages 153--158. IEEE, 2000.

\bibitem{athar2006report}
MS~Athar, INO Collaboration, et~al.
\newblock a report of the ino feasibility study.
\newblock {\em Updated from the earlier Interim Report of May}, 1:2005, 2006.

\bibitem{bheesette2009design}
Satyanarayana Bheesette.
\newblock {\em Design and Characterisation Studies of Resistive Plate
  Chambers}.
\newblock PhD thesis, INDIAN INSTITUTE OF TECHNOLOGY BOMBAY, 2009.

\bibitem{Agostinelli2003250}
S.~Agostinelli et~al.
\newblock Geant4 - a simulation toolkit.
\newblock {\em Nuclear Instruments and Methods in Physics Research Section A:
  Accelerators, Spectrometers, Detectors and Associated Equipment}, 506(3):250
  -- 303, 2003.

\bibitem{devi2013hadron}
Moon~Moon Devi, Anushree Ghosh, Daljeet Kaur, SM~Lakshmi, Sandhya Choubey, Amol
  Dighe, D~Indumathi, Sanjeev Kumar, MVN Murthy, and Md~Naimuddin.
\newblock Hadron energy response of the iron calorimeter detector at the
  india-based neutrino observatory.
\newblock {\em Journal of Instrumentation}, 8(11):P11003, 2013.

\bibitem{Ghosh:2009ea}
Tapasi Ghosh and Subhasis Chattopadhyay.
\newblock {Track fitting by Kalman Filter method for a prototype cosmic ray
  muon detector}.
\newblock 2009.

\bibitem{MuLkUpTbl}
Animesh Chatterji, Kanishka Rawat, Meghna, and Tarak Thakore.
\newblock Muon reconstruction look up table, v2.
\newblock Technical report, INO Collaboration, 2012.

\bibitem{9780198508717}
Carlo Giunti and Chung~W. Kim.
\newblock {\em Fundamentals of Neutrino Physics and Astrophysics}.
\newblock Oxford University Press, USA, 2007.

\bibitem{smirnov2013neutrino}
Alexei~Yu Smirnov.
\newblock Neutrino mass hierarchy and matter effects.
\newblock 2013.

\bibitem{Samanta:2006sj}
Abhijit Samanta.
\newblock {The Mass hierarchy with atmospheric neutrinos at INO}.
\newblock {\em Phys.Lett.}, B673:37--46, 2009.

\bibitem{Samanta:2007ue}
Abhijit Samanta, Sudeb Bhattacharya, Ambar Ghosal, Kamales Kar, Debasish
  Majumdar, et~al.
\newblock {A GEANT-based study of atmospheric neutrino oscillation parameters
  at INO}.
\newblock {\em Int.J.Mod.Phys.}, A23:233--245, 2008.

\bibitem{Ghosh:2012px}
Anushree Ghosh, Tarak Thakore, and Sandhya Choubey.
\newblock {Determining the Neutrino Mass Hierarchy with INO, T2K, NOvA and
  Reactor Experiments}.
\newblock {\em JHEP}, 1304:009, 2013.

\bibitem{Thakore:2013xqa}
Tarak Thakore, Anushree Ghosh, Sandhya Choubey, and Amol Dighe.
\newblock {The Reach of INO for Atmospheric Neutrino Oscillation Parameters}.
\newblock {\em JHEP}, 1305:058, 2013.

\bibitem{Fruhwirth:1987fm}
R.~Fr{\"u}hwirth.
\newblock {Application of Kalman filtering to track and vertex fitting}.
\newblock {\em Nucl.Instrum.Meth.}, A262:444--450, 1987.

\bibitem{fontana2007track}
A~Fontana, P~Genova, L~Lavezzi, and A~Rotondi.
\newblock Track following in dense media and inhomogeneous magnetic fields.
\newblock {\em PANDA Report PVI01-07}, pages pp. 3--6, 19--28, 2007.

\bibitem{Gorbunov:2006pe}
S.~Gorbunov and I.~Kisel.
\newblock {Analytic formula for track extrapolation in non-homogeneous magnetic
  field}.
\newblock {\em Nucl.Instrum.Meth.}, A559:148--152, 2006.

\bibitem{9783540327950}
Juan~Andrade Cetto and Alberto Sanfeliu.
\newblock {\em Environment Learning for Indoor Mobile Robots: A Stochastic
  State Estimation Approach to Simultaneous Localization and Map Building
  (Springer Tracts in Advanced Robotics)}.
\newblock Springer, 2006.

\bibitem{Nakamura:2010zzi}
K.~Nakamura et~al.
\newblock {Review of particle physics}.
\newblock {\em J.Phys.}, G37:075021, 2010.

\bibitem{Wittek:931170}
W~Wittek.
\newblock {Propagation of errors along a particle trajectory in a magnetic
  field}.
\newblock Technical Report EMC-80-15, EMC-80-15-ADD-1, CERN, Geneva, 1980-81.

\bibitem{mathematica2011inc}
Wolfram~Res Mathematica.
\newblock Inc., champaigne il, 2011.

\bibitem{9783540572800}
William~R. Leo.
\newblock {\em Techniques for Nuclear and Particle Physics Experiments: A
  How-to Approach}.
\newblock Springer, 1994.

\bibitem{Wittek:931172}
W~Wittek.
\newblock {Error propagation with energy loss: second addendum. Propagation of
  errors along a particle trajectory in a magnetic field}.
\newblock Technical Report EMC-80-15-ADD-2, CERN, Geneva, May 1982.

\bibitem{Groom:2001kq}
Donald~E. Groom, Nikolai~V. Mokhov, and Sergei~I. Striganov.
\newblock {Muon stopping power and range tables 10-MeV to 100-TeV}.
\newblock {\em Atom.Data Nucl.Data Tabl.}, 78:183--356, 2001.

\bibitem{fruhwirth2001quantitative}
R~Fr{\"u}hwirth and M~Regler.
\newblock On the quantitative modelling of core and tails of multiple
  scattering by gaussian mixtures.
\newblock {\em Nuclear Instruments and Methods in Physics Research Section A:
  Accelerators, Spectrometers, Detectors and Associated Equipment},
  456(3):369--389, 2001.

\bibitem{lynch1991approximations}
Gerald~R Lynch and Orin~I Dahl.
\newblock Approximations to multiple coulomb scattering.
\newblock {\em Nuclear Instruments and Methods in Physics Research Section B:
  Beam Interactions with Materials and Atoms}, 58(1):6--10, 1991.

\bibitem{lassila1995energy}
Kati Lassila-Perini and L{\'a}szl{\'o} Urb{\'a}n.
\newblock Energy loss in thin layers in geant.
\newblock {\em Nuclear Instruments and Methods in Physics Research Section A:
  Accelerators, Spectrometers, Detectors and Associated Equipment},
  362(2):416--422, 1995.

\bibitem{hoppner2010novel}
C~H{\"o}ppner, S~Neubert, B~Ketzer, and S~Paul.
\newblock A novel generic framework for track fitting in complex detector
  systems.
\newblock {\em Nuclear Instruments and Methods in Physics Research Section A:
  Accelerators, Spectrometers, Detectors and Associated Equipment},
  620(2):518--525, 2010.

\bibitem{marshall2008study}
John~Stuart Marshall.
\newblock {\em A study of muon neutrino disappearance with the MINOS detectors
  and the NuMI neutrino beam}.
\newblock PhD thesis, University of Cambridge, 2008.

\bibitem{strandlie2006derivation}
A~Strandlie and W~Wittek.
\newblock Derivation of jacobians for the propagation of covariance matrices of
  track parameters in homogeneous magnetic fields.
\newblock {\em Nuclear Instruments and Methods in Physics Research Section A:
  Accelerators, Spectrometers, Detectors and Associated Equipment},
  566(2):687--698, 2006.

\end{thebibliography}
\end{document}